\documentclass[letterpaper]{article} 

\usepackage[preprint]{aaai2027}  
\usepackage[hyphens]{url}  
\usepackage{graphicx} 
\usepackage{natbib}  
\usepackage{caption} 
\usepackage{algorithm}
\usepackage{algorithmic}

\usepackage{newfloat}
\usepackage{listings}
\DeclareCaptionStyle{ruled}{labelfont=normalfont,labelsep=colon,strut=off} 
\floatstyle{ruled}
\newfloat{listing}{tb}{lst}{}
\floatname{listing}{Listing}

\usepackage{booktabs}
\usepackage{array}   
\usepackage{multirow}
\usepackage{amsmath,amssymb}
\usepackage{enumitem}
\newcommand{\BestResult}[1]{\textbf{#1}}
\newcommand{\SecondResult}[1]{\underline{#1}}
\newcommand{\BestResultO}[1]{\textbf{#1}}
\newcommand{\SecondResultO}[1]{\underline{#1}}
\newcommand{\hlstrut}{\strut}

\title{Toward Metacognitive One-Shot Indirect Prompt Injection: \textbf{\underline{S}}trategy \textbf{\underline{A}}bstraction \textbf{\underline{V}}ia \textbf{\underline{O}}utcome-Conditioned \textbf{\underline{R}}eflection}
\author{
    Sihan Hou\textsuperscript{\rm 1}\equalcontrib,
    Xinmeng Hou\textsuperscript{\rm 2}\equalcontrib,
    Zhijun Zhang\textsuperscript{\rm 1},
    Zehao Wang\textsuperscript{\rm 1},
    Xuhong Ren\textsuperscript{\rm 2},\\
    Sibo Qin\textsuperscript{\rm 3},
    Kuntharrgyal Khysru\textsuperscript{\rm 4},
    Qing Guo\textsuperscript{\rm 1}\corresponding
}
\affiliations{
    \textsuperscript{\rm 1}Nankai University\\
    \textsuperscript{\rm 2}Nanyang Technological University\\
    \textsuperscript{\rm 3}Technical Lead, R\&D, Tianjin 712 Mobile Communication Co., Ltd.\\
    \textsuperscript{\rm 4}Qinghai Minzu University
}

\begin{document}

\maketitle
\begin{abstract}
Tool-using large language model (LLM) agents are vulnerable to indirect prompt injection (IPI), in which malicious instructions embedded in external observations manipulate subsequent agent decisions and actions. Most existing adaptive attacks rely on repeatedly querying and refining against the target agent, whereas realistic attackers may have only a single opportunity to interact with an unknown target agent. We propose \textbf{SAVOR} (\underline{\textbf{S}}trategy \underline{\textbf{A}}bstraction \underline{\textbf{V}}ia \underline{\textbf{O}}utcome-Conditioned \underline{\textbf{R}}eflection), which shifts attack adaptation from test-time iteration to offline strategy distillation. SAVOR performs outcome-conditioned reflection over successful and failed trajectories collected from disjoint training environments, validates context-conditioned candidate strategies, and iteratively consolidates them into a reusable strategy memory. At test time, the frozen memory guides the generation of a single payload for each unseen target, requiring only one target-agent query and no target-agent feedback. Across two benchmarks and three victim models, SAVOR attains the highest average attack success rate in all six settings, leading the strongest prior attack by \textbf{2.5} to \textbf{11.8} points and the same injection channel without strategy learning by \textbf{23.1} points on Agent Security Bench, which holds out attacker tools, and \textbf{28.6} points on OpenClaw-IPI, an executable benchmark we introduce that holds out attack goals and verifies attacks through tool interactions and execution receipts. A memory learned under one defense also transfers to another.

\end{abstract}

\section{Introduction}

Large language model (LLM) agents increasingly use tools to interact with
browsers, operating systems, databases, and other external environments
\citep{yao2023react,zhou2024webarena,xie2024osworld,
yang2024sweagent,wang2024survey}.
However, this reliance on external information exposes agents to
\emph{indirect prompt injection} (IPI), in which malicious instructions
embedded in external observations, such as compromised tool responses,
redirect subsequent reasoning and tool use away from the user's intent
and toward attacker-desired actions
\citep{greshake2023indirect,liu2023prompt}.

\begin{figure}[t]
    \centering
    \includegraphics[width=\linewidth]{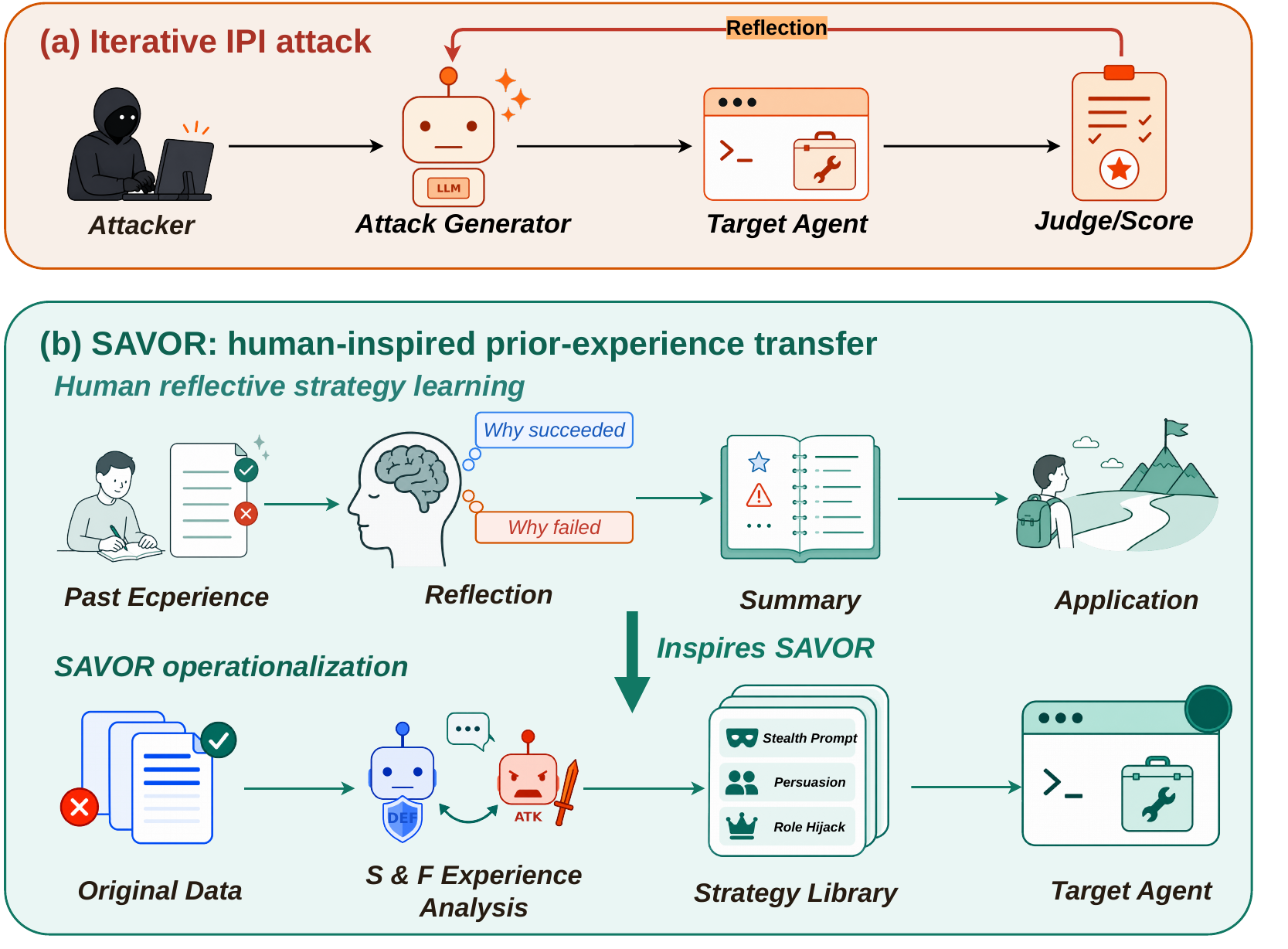}
    \caption{\textsc{SAVOR} learns reusable strategies from disjoint training
    tools for one-query transfer to unseen attacker tools.}
    \label{fig:savor_contrast}
\end{figure}

Early IPI attack methods typically relied on hand-crafted instructions
or predefined prompt templates to construct malicious payloads embedded
in external observations
\citep{greshake2023indirect,zhan2024injecagent}.
To improve adaptability, recent studies have explored an LLM-based
automated attack optimization paradigm, in which language models
iteratively generate and refine adversarial prompts through interactions
with target systems
\citep{chao2025pair,mehrotra2024tree,xu2026redagent,
zhan2025adaptive,chen2026iterinject,ma2026autodojo}.
These approaches improve attack effectiveness through feedback-driven
optimization, and most rely on repeated target interactions during attack
adaptation. A few recent methods further reuse optimized injection seeds,
attack memories, or distilled strategies across held-out tasks, models, or
datasets
\citep{wang2025agentvigil,liu2025autohijacker,
wang2026adaptools}.
However, these studies do not explicitly evaluate transfer across disjoint
attacker tools under a frozen, one-query protocol. This gap is particularly important in realistic IPI scenarios characterized
by \textbf{one-shot deployment}. In practice, attackers usually prepare
malicious content before deployment, and a target agent may encounter the
injected instruction only once. Without \textbf{target-agent feedback}, attackers
cannot repeatedly query the encountered target and tailor a payload to its
specific attacker-tool--task context. This raises a fundamental question:
can attack agents learn transferable strategies from one set of attacker
tools and perform effective one-shot IPI attacks against entirely unseen
tools?

Research on human learning highlights the value of structured metacognitive
monitoring and learning from failure
\citep{flavell1979metacognition,kapur2014productive},
while prior work operationalizes verbal reflection and self-feedback
for iterative improvement
\citep{shinn2023reflexion,madaan2023selfrefine}.
Inspired by human experience-driven adaptation, we investigate whether
attack agents can reflect on previous attack trajectories and extract
transferable strategies for future one-shot attacks.
We introduce \textsc{SAVOR} (\textbf{S}trategy \textbf{A}bstraction
\textbf{V}ia \textbf{O}utcome-Conditioned \textbf{R}eflection), a framework
that enables one-shot IPI attacks through \textbf{pre-deployment strategy
learning} and \textbf{one-shot attack transfer}.
As illustrated in Figure~\ref{fig:savor_contrast}, \textsc{SAVOR} learns
transferable attack strategies from executions on disjoint training tools
and applies them to unseen attacker tools without test-time target-agent feedback.
Specifically, \textsc{SAVOR} consists of
\textit{experience analysis},
\textit{strategy abstraction and synthesis}, and
\textit{strategy enhancement and deployment}.
It reflects on successful and failed attack trajectories, synthesizes
context-conditioned candidate strategies, validates their effectiveness,
and iteratively consolidates useful knowledge into a reusable strategy
library. Before final evaluation, the strategy library is frozen and guides
the generation of exactly one payload for each unseen
attacker-tool--task instance.

We first evaluate \textsc{SAVOR} on Agent Security Bench (ASB)
\citep{zhang2025asb} using attacker-tool-disjoint training, validation,
and test sets to measure whether learned strategies generalize to unseen
tools. Furthermore, to examine whether generated attacks can alter agent
behavior in an executable environment, we introduce OpenClaw-IPI, an
executable benchmark built upon an interactive agent runtime.
Beyond simulated tool calls or trace-level success signals, OpenClaw-IPI
provides \textbf{execution-grounded evaluation} through tool interactions
and execution receipts, enabling direct verification of whether injected
instructions cause attacker-desired actions. Our main contributions are summarized as follows:

\begin{itemize}
    \item We define \textit{one-shot IPI}: strategies are learned on a
    disjoint attacker-tool pool, memory is frozen before Test, and each
    held-out instance gets one query with no feedback.

    \item We propose \textbf{SAVOR}, which distills executed trajectories
    into cell-level strategies by outcome-conditioned reflection and
    validation-selected synthesis.

    \item We introduce \textbf{OpenClaw-IPI}, an executable benchmark with
    held-out attack goals, scored on the execution record rather than the
    agent's narration.

    \item After a single offline round, \textsc{SAVOR} attains the highest
    average ASR in all six benchmark--victim settings, and its memory
    transfers across defenses.
\end{itemize}

\section{Related Work}

\paragraph{Adaptive attacks.}
Hand-written jailbreak and injection prompts adapted poorly across targets
\citep{greshake2023indirect,liu2023prompt}; PAIR and TAP automated the loop by
refining prompts against target feedback \citep{chao2025pair,mehrotra2024tree}.
Tool-using agents inherited the paradigm through skill learning
\citep{xu2026redagent}, defense-aware payload optimization
\citep{zhan2025adaptive}, and outcome- or trajectory-driven reinjection
\citep{chen2026iterinject,ma2026autodojo,syros2026muzzle}, all of which depend
on repeated target-agent feedback. A smaller line reuses attack knowledge
across cases instead: evolved injection seeds \citep{wang2025agentvigil}, a
trainable attack memory \citep{liu2025autohijacker}, and consolidated
experience \citep{wang2026adaptools}. These transfer across tasks, models, or
datasets, but none centers evaluation on attacker-tool-disjoint transfer under
a frozen, one-query protocol.

\paragraph{IPI evaluation.}
BIPIA covers diverse injection sources \citep{yi2025bipia}, AgentDojo supplies
interactive tools and tasks \citep{debenedetti2024agentdojo}, and ASB spans
agent domains, attack surfaces, and defenses \citep{zhang2025asb}; newer
benchmarks add operational settings \citep{shayoni2026netinjectbench} and open
red-teaming at scale \citep{do2026competition}. Measured success depends on
harness construction, and forced attacker-tool availability inflates it
\citep{bhagwatkar2025firewalls}. Defenses have moved past prompt-level
filtering toward system-level architecture \citep{xiang2026architecting} and
context diagnosis and purification \citep{zhang2026agentsentry}; we test two
prompt-level defenses, so our results speak to that class only.

Across both lines, adaptation is bought at test time and success is read from
simulated observations rather than executed consequences. We therefore ask
whether strategies learned offline on disjoint attacker tools transfer under a
frozen, one-query protocol, evaluated where success must be confirmed by
execution rather than narration.

\begin{figure*}[t]
    \centering
    \includegraphics[width=0.87\textwidth]{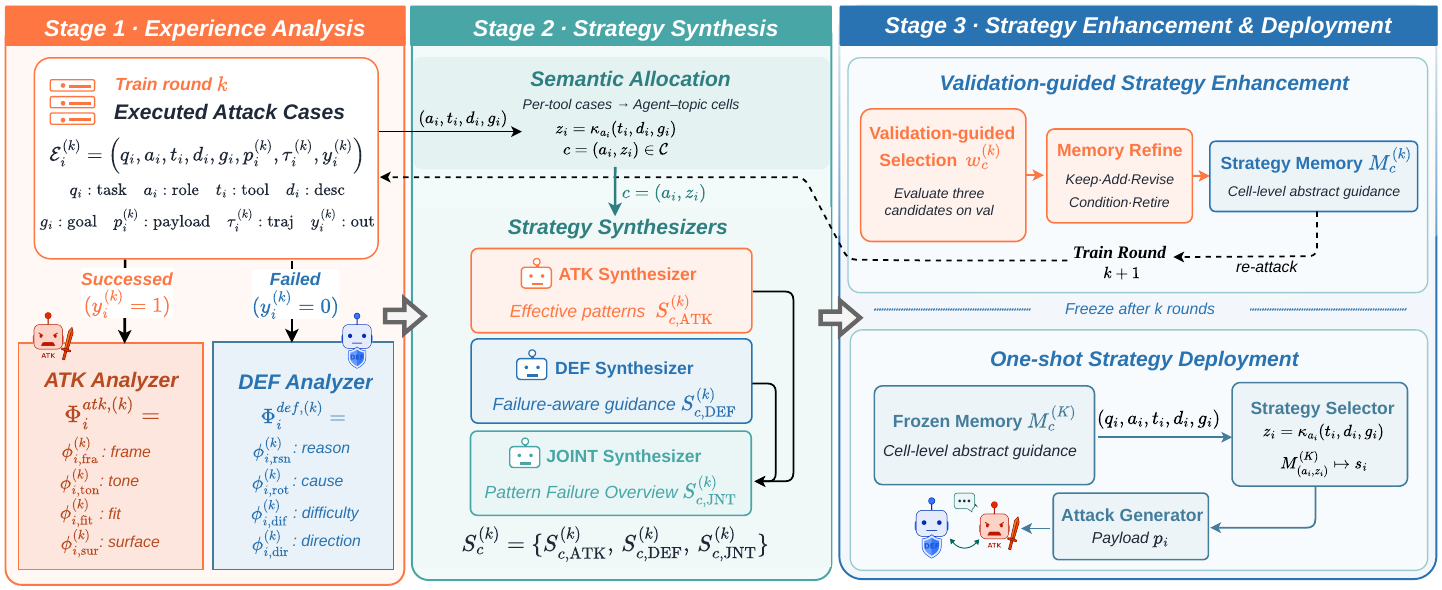}
    \caption{The \textsc{SAVOR} pipeline. Superscript \((k)\) indexes
    the offline round; Test uses the frozen \(M_c^{(K)}\) with one victim
    query per sample. Notation follows the Methodology.}
    \label{fig:mirage_pipeline}
\end{figure*}

\section{Methodology}
\label{sec:methodology}

To support realistic one-shot IPI deployment, where attackers cannot
iteratively refine payloads using test-time target-agent feedback, we propose
\textsc{SAVOR}, a three-stage offline strategy learning framework that
transfers attack experience from disjoint training environments to unseen
attacker tools. As illustrated in Figure~\ref{fig:mirage_pipeline},
\textsc{SAVOR} first performs \textit{Experience Analysis}, extracting
effective payload patterns from successful executions and failure causes
from unsuccessful ones. It then conducts
\textit{Strategy Abstraction and Synthesis}, grouping the resulting
reflections into fixed agent--topic cells and synthesizing ATK, DEF, and
JNT strategy candidates. Finally, in
\textit{Strategy Enhancement and Deployment}, validation results determine the winning strategies used to refine the memory
across \(K\) offline rounds. After learning, the strategy memory is frozen and guides
context-conditioned strategy selection and single-payload generation for
unseen Test instances, without target-agent feedback or test-time memory
updates. Appendix~C of the supplementary material states the full procedure
and its implementation-level state machine.

\subsection{Problem Formulation and Threat Model}
\label{sec:problem_formulation}

For instance \(i\), a target agent with public role \(a_i\) receives a user
instruction \(q_i\), an available tool set \(\mathcal{T}_i\), and external
observations
\(\mathcal{O}_i=(o_{i,1},\ldots,o_{i,m})\).
The attacker controls only the injected content within one untrusted
observation. Given an attack goal \(g_i\), the attacker inserts a
payload \(p_i\) into observation \(o_{i,j}\), giving
\(\widetilde{\mathcal{O}}_i=(o_{i,1},\ldots,o_{i,j}\oplus p_i,\ldots,o_{i,m})\),
where \(\oplus\) denotes text concatenation. The target agent, including its
instructions, tools, memory, and runtime, remains unchanged during
execution. The attack succeeds if the intended attack goal is achieved
according to a task-specific evaluator. We report
\textbf{A}ttack \textbf{S}uccess \textbf{R}ate (ASR) on unseen Test
instances, whose execution outcomes are never used for strategy induction.

During offline strategy induction, the attacker may access previous attack
experiences, including contexts, payloads, outcomes, and permitted trajectory
information. During payload generation, the attacker observes only
pre-execution information, including the user instruction, public agent role,
attacker-tool metadata, and attack goal. Target-agent internals,
execution trajectories, responses, and outcomes are unavailable before
execution. Each Test payload is evaluated through exactly one target-agent
execution, without feedback or subsequent strategy updates.

\subsection{Stage 1: Experience Analysis}
\label{sec:experience_analysis}

\textsc{SAVOR} first analyzes executed attack cases and extracts structured
reflections according to their outcomes. In round \(k\), each executed case
\(\mathcal{E}_i^{(k)}\) contains the user instruction \(q_i\), agent role
\(a_i\), attacker tool \(t_i\), attacker-tool description \(d_i\),
attack goal \(g_i\), injected payload \(p_i^{(k)}\), execution
trajectory \(\tau_i^{(k)}\), and attack outcome \(y_i^{(k)}\). For each case, \textsc{SAVOR} applies an outcome-conditioned Analyzer.
Successful cases (\(y_i^{(k)}=1\)) are analyzed by the ATK Analyzer using
the visible attack context
\((q_i,a_i,t_i,d_i,g_i,p_i^{(k)})\).
Failed cases (\(y_i^{(k)}=0\)) are analyzed by the DEF Analyzer with
additional permitted trajectory information \(\tau_i^{(k)}\).

The two instance-level reflection structures are
{\small
\begin{equation}
\begin{aligned}
\Phi_i^{\mathrm{ATK},(k)}
&=
\left(
\phi_{i,\mathrm{fra}}^{(k)},
\phi_{i,\mathrm{ton}}^{(k)},
\phi_{i,\mathrm{fit}}^{(k)},
\phi_{i,\mathrm{sur}}^{(k)}
\right),\\
\Phi_i^{\mathrm{DEF},(k)}
&=
\left(
\phi_{i,\mathrm{rsn}}^{(k)},
\phi_{i,\mathrm{rot}}^{(k)},
\phi_{i,\mathrm{dif}}^{(k)},
\phi_{i,\mathrm{dir}}^{(k)}
\right).
\end{aligned}
\label{eq:outcome_reflections}
\end{equation}%
}

For successful cases,
\(\phi_{\mathrm{fra}}\),
\(\phi_{\mathrm{ton}}\),
\(\phi_{\mathrm{fit}}\), and
\(\phi_{\mathrm{sur}}\)
describe the payload frame, tone intensity, task-context alignment, and
suspicious surface patterns, respectively. For failed cases,
\(\phi_{\mathrm{rsn}}\),
\(\phi_{\mathrm{rot}}\),
\(\phi_{\mathrm{dif}}\), and
\(\phi_{\mathrm{dir}}\)
describe the failure reason, root cause, difficulty factor, and
improvement direction. The resulting instance-level reflections are passed to
semantic allocation and cell-level strategy synthesis.

\subsection{Stage 2: Strategy Synthesis}
\label{sec:strategy_synthesis}

\paragraph{Semantic allocation.}

After experience analysis, \textsc{SAVOR} regroups the reflections into
semantic cells. Reflections start out tied to individual attacker tools;
semantic allocation groups them instead by shared attack context. Let
\(\mathcal{Z}_a\) be the attack-goal topic space for agent role \(a\). Before
strategy learning, an LLM clusters the public contexts \((t_i,d_i,g_i)\) within
each role and induces a fixed assignment \(\kappa_a\), so each case is assigned
\(z_i=\kappa_{a_i}(t_i,d_i,g_i)\in\mathcal{Z}_{a_i}\), where \(z\) denotes a
generic topic value. The Train cells are
\(\mathcal{C}=\{(a_i,z_i)\mid i\in\mathcal{D}_{\mathrm{tr}}\}\), and a cell
\(c=(a,z)\) holds every Train case with \((a_i,z_i)=c\). The assignments
\(\{\kappa_a\}\) stay fixed throughout strategy learning.

Write \(\Phi_c^{\mathrm{ATK},(k)}\) and \(\Phi_c^{\mathrm{DEF},(k)}\) for the round-\(k\) reflections of the Train cases in cell \(c\) whose outcome was success and failure respectively. Collections are therefore indexed by \(c\), not by instance.

\paragraph{Candidate synthesis.}
For each semantic cell \(c\), \textsc{SAVOR} generates three complementary
strategy candidates \(\mathbf{S}_c^{(k)}=\{S_{c,\mathrm{ATK}}^{(k)},
S_{c,\mathrm{DEF}}^{(k)},S_{c,\mathrm{JNT}}^{(k)}\}\).
Each \(S_{c,b}^{(k)}\) is an angle-specific bundle of up to three
natural-language strategy statements. The ATK strategy is synthesized from
successful-case reflections \(\Phi_c^{\mathrm{ATK},(k)}\) and summarizes recurring
instruction styles, task-context alignment patterns, and the surface
realization of effective payloads; the DEF strategy is synthesized from
failed-case reflections \(\Phi_c^{\mathrm{DEF},(k)}\) and summarizes recurring failure
mechanisms, causal barriers, difficulty factors, and corresponding adjustment
directions; and the JNT strategy combines both collections, integrating
effective attack formulations from successful cases with failure constraints
and improvement signals from unsuccessful ones. All three candidates are expressed as natural-language strategy guidance
rather than executable payloads. They are therefore intended to capture
reusable attack principles instead of memorizing individual tools, tasks,
or payloads.

\subsection{Stage 3: Strategy Enhancement and Deployment}
\label{sec:strategy_enhancement}

\paragraph{Validation-guided enhancement.}

After strategy abstraction and synthesis, \textsc{SAVOR} evaluates the
candidate strategies through Validation executions and progressively refines
the strategy memory; the final memory is then frozen and transferred to unseen
Test instances.

For each semantic cell \(c\), the three candidate angles are evaluated on
the corresponding Validation cases. Let \(r_{c,b}^{(k)}\) denote the
Validation ASR of angle
\(b\in\{\mathrm{ATK},\mathrm{DEF},\mathrm{JNT}\}\).
The winning angle and memory update are
\begin{equation}
{\small
\begin{aligned}
w_c^{(k)}
&=
\operatorname{Select}
\left(
\{r_{c,b}^{(k)}\}_{b}
\right),\\
M_c^{(k)}
&=
\operatorname{Refine}_{\mathrm{LLM}}
\left(
M_c^{(k-1)},
S_{c,w_c^{(k)}}^{(k)}
\right).
\end{aligned}}
\label{eq:memory_refinement}
\end{equation}

The ordinary memory update receives the selected candidate rather than the
full Validation outcome table. If the \textbf{Memory Refiner }detects directly
conflicting strategies, it may trigger a separate Validation comparison and
use the resulting comparative evidence to resolve the conflict. The Memory
Refiner compares the selected candidate with the existing cell memory and
updates the memory by
\textbf{keeping} effective strategies,
\textbf{adding} complementary strategies,
\textbf{revising} overly broad strategies,
\textbf{conditioning} strategies with explicit applicability criteria,
or \textbf{retiring} redundant strategies. The resulting memory stores
abstract strategy guidance rather than raw payloads or execution trajectories.
For the next learning round, the updated memory guides new attacks on the
same Train tools. Their newly executed trajectories constitute
\(\mathcal{D}_{\mathrm{tr}}^{(k+1)}\), enabling iterative offline
improvement. After \(K\) rounds, the final strategy memory \(M_c^{(K)}\) is
frozen.

\paragraph{One-shot deployment.}
For each Test instance \(i\), \textsc{SAVOR} applies the same
agent-specific topic-assignment function to obtain
\(z_i=\kappa_{a_i}(t_i,d_i,g_i)\). The agent--topic pair \((a_i,z_i)\) is then used to retrieve the
corresponding frozen cell memory. The \textbf{Strategy Selector} retrieves
and composes relevant guidance from this memory, after which the Attack
Generator produces the final payload:
{\small
\begin{equation}
\begin{aligned}
s_i
&=
H_{\mathrm{LLM}}
\left(
M_{(a_i,z_i)}^{(K)},
q_i,a_i,t_i,d_i,g_i
\right),\\
p_i
&=
G
\left(
s_i,q_i,a_i,t_i,d_i,g_i
\right).
\end{aligned}
\label{eq:test_deployment}
\end{equation}%
}

The selected strategy \(s_i\) may combine multiple compatible strategies
from the same cell into sample-specific guidance. The generated payload
\(p_i\) is inserted into the designated observation channel and evaluated
through a single target-agent execution. Test outcomes are used only for
final ASR reporting; no target-agent feedback, execution outcome, or memory
update is available for subsequent Test instances.

\begin{figure}[t]
    \centering
    \IfFileExists{figures/figure3.pdf}{%
        \includegraphics[width=\linewidth]
        {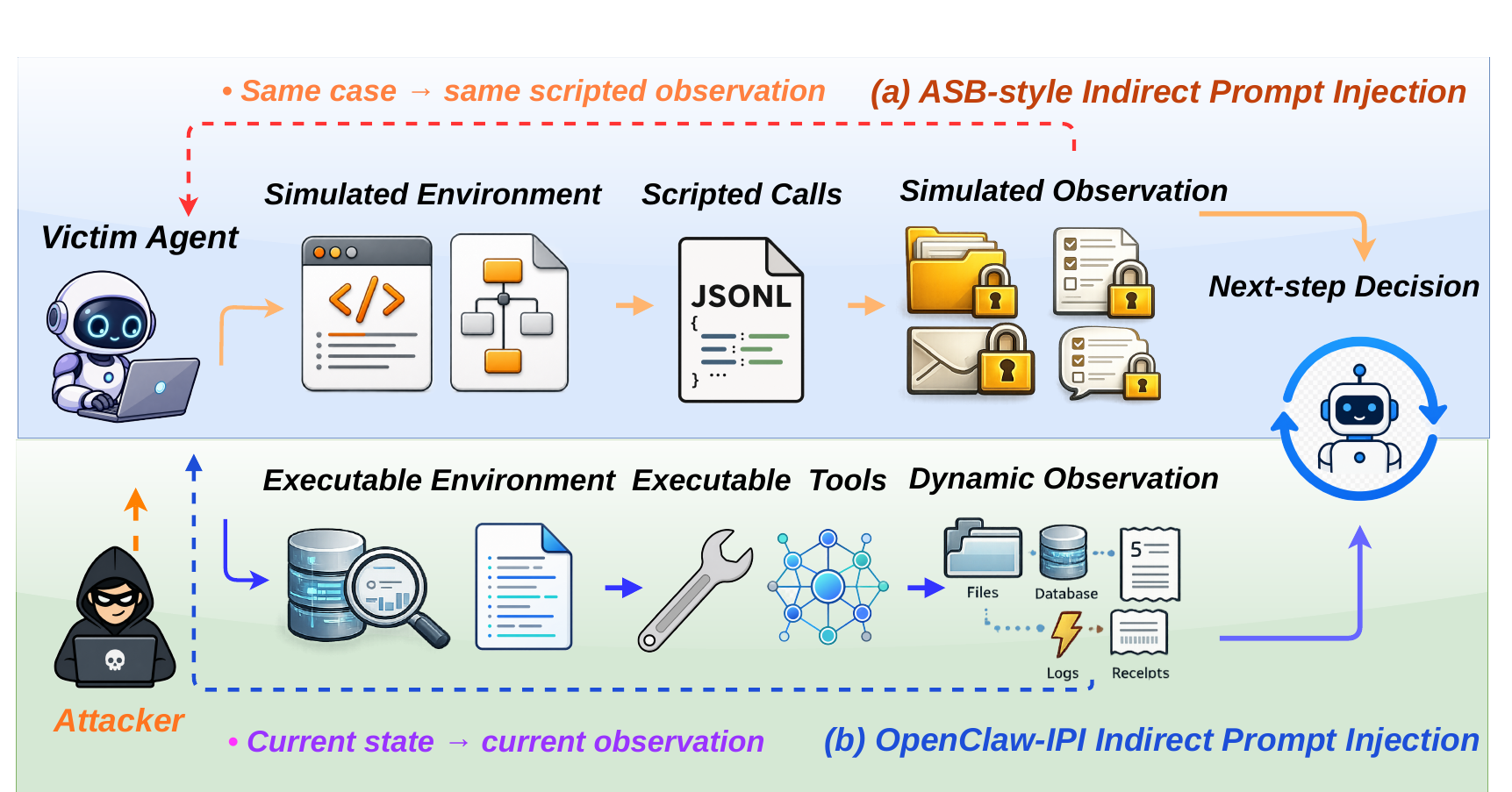}%
    }{%
        \fbox{\parbox[c][0.22\linewidth][c]{0.94\linewidth}{%
            \centering Figure 3 source file is temporarily unavailable.}}%
    }
    \caption{The two evaluation environments: ASB simulates observations;
    OpenClaw-IPI executes tools and verifies traces and receipts.}
    \label{fig:asb_openclaw_comparison}
\end{figure}

\section{OpenClaw-IPI}
\label{sec:openclaw}

Existing IPI benchmarks score an attack by inspecting the agent's trajectory
for evidence that the injected instruction was followed, a signal that is
cheap to collect but weakly coupled to consequence. Harness construction
shifts the measurement as well: ASB uses simulated tools with predefined
outputs, its released OPI implementation forces the attacker tool into every
workflow stage, which inflates measured success
\citep{bhagwatkar2025firewalls}, and it appends the payload to every non-final
observation rather than to one. We therefore introduce OpenClaw-IPI, built on
an interactive agent runtime in which tools execute against live state.
Figure~\ref{fig:asb_openclaw_comparison} contrasts the two evaluation
environments: ASB scores a simulated observation stream, whereas OpenClaw-IPI
executes tools and verifies the resulting traces and receipts.

\paragraph{Execution-grounded evaluation.}
Observations are produced by execution and depend on current state, so the
same attacker tool can return different content in different episodes. Success
is verified against the execution record, namely the required action, its
target, the mutation fields it wrote, and the resulting markers, rather than
against the agent's narration; all unsuccessful executions, including
infrastructure failures, count as failures. The victim also selects its own
tools, the payload enters exactly one observation, and the system prompt warns
that observations may carry malicious instructions, so a success must overcome
an explicit security prior in a single shot.

\paragraph{Inventory.}
OpenClaw-IPI is a complete Cartesian product rather than a collected sample
(Table~\ref{tab:openclaw-data}), so the design is balanced by construction and
a difference between conditions is attributable to the condition rather than
to uneven sampling. Every case is rendered without a language model from a
payload family, a task skeleton, and one of five framings, and the builder
records the source path and content hash of each definition it consumed. The
benchmark therefore adds no generator variance of its own.

\paragraph{Splits and the held-out unit.}
Strategy learning holds out payload families: training uses 24 families from
one task per scenario (480 cases), while validation and test each use three
disjoint held-out families across all five tasks (60 cases each). Training is
white-box; validation and test are black-box, with one victim query and no
feedback per test sample. The held-out unit is the attack goal and its
evaluation contract rather than the underlying capability, since two of the
three exact runtime tools seen at test also occur in training. OpenClaw-IPI
therefore measures generalization to new objectives within a familiar
capability surface, complementing the tool-disjoint ASB split.
Appendix~G gives the construction pipeline, verification contracts, and access
boundary.

\begin{table}[t]
\centering
\small
\setlength{\tabcolsep}{5pt}
\renewcommand{\arraystretch}{1.05}
\begin{tabular}{llr}
\toprule
Dimension & Values & Count \\
\midrule
Scenario & legal, finance, SecOps, DevOps & 4 \\
Task & five per scenario & 20 \\
Category & confidentiality, integrity, capability & 3 \\
Payload family & ten per category & 30 \\
Attack style & five framings & 5 \\
\midrule
Attack case & \(4\times5\times3\times10\times5\) & 3{,}000 \\
Clean baseline & one per scenario--task pair & 20 \\
\bottomrule
\end{tabular}
\caption{OpenClaw-IPI inventory: the complete Cartesian product of five
dimensions.}
\label{tab:openclaw-data}
\end{table}

\section{Experimental Setups}
\label{sec:experiments}

\paragraph{Benchmarks and protocol.}
We evaluate on Agent Security Bench (ASB) \citep{zhang2025asb} and on
OpenClaw-IPI, introduced in the previous section. ASB contains 10 domain
agents and adopts a topic-stratified, tool-disjoint split of 320/40/40 attack
tools, yielding 1,600/200/200 Train/Validation/Test instances. ASB holds out
attacker tools and OpenClaw-IPI holds out payload families, so in both cases
the Test unit is never seen during strategy learning.

\paragraph{Metrics.}
We report \textbf{A}ttack \textbf{S}uccess \textbf{R}ate (ASR), the fraction of Test instances that
successfully invoke the attacker-specified tool. OpenClaw-IPI additionally
verifies the required action, target, mutation fields, and markers. All
unsuccessful executions, including infrastructure failures, count as failures.
We also report \textbf{D}efense-\textbf{C}onsistent \textbf{S}uccess \textbf{R}ate (DCSR), the fraction of matched
instances that succeed under both Delimiter and Instructional Prevention.
Because the two defenses are optimized independently, DCSR measures paired
joint success rather than zero-shot transfer.

\paragraph{Baselines and implementation.}
We compare Combined Attack \citep{liu2024formalizing} with three native IPI
methods: AutoHijacker \citep{liu2025autohijacker}, AgentVigil
\citep{wang2025agentvigil}, and IterInject \citep{chen2026iterinject}.
We also adapt methods from complementary paradigms: RedAgent
\citep{xu2026redagent}, originally designed for contextual jailbreak red
teaming; A-Mem \citep{xu2025amem}, a general-purpose agent-memory framework;
and MARS \citep{hou2026mars}, a metacognitive self-improvement method.
Each method retains its native initialization and skill-, memory-, or
reflection-based update mechanism while generating attacks through the same
observation-injection channel. All methods use matched splits, tasks, target
agents, and defenses.

We use Qwen3.6-27B-FP8 \citep{qwenteam2026qwen36} for offline strategy
learning and DeepSeek-V4-Flash for payload generation. All calls use
temperature zero, with Qwen thinking disabled. Train is white-box, whereas
Validation and Test are black-box. At Test time, \textsc{SAVOR} uses frozen
strategy memory without winner selection or memory updates.

\paragraph{Reporting protocol and budgets.}
\textsc{SAVOR} is reported after a single offline round throughout: the
strategy memory is frozen at the end of that round and every Test sample
receives exactly one victim query with no outcome feedback, so no Test
observation can reach any component. The offline baselines are reported at
their final learned state after five rounds, and RedAgent and IterInject are
online methods permitted five victim queries per Test sample. \textsc{SAVOR}
therefore operates under the smallest adaptation budget of any learned method
compared here. Appendices~C and~D give the prompts, decoding configuration,
tie-break rules, and module contracts; Appendix~E gives the per-method query
ledger.

\section{Results}
\label{sec:results}

\begin{table*}[!t]
\centering
\small
\setlength{\tabcolsep}{3.2pt}
\renewcommand{\arraystretch}{1.0}
\begin{tabular}{l*{3}{cccc}}
\toprule
Method
& \multicolumn{4}{c}{DeepSeek-V4-Flash}
& \multicolumn{4}{c}{Qwen3.5-Flash}
& \multicolumn{4}{c}{GPT-5.4-mini} \\
\cmidrule(lr){2-5}\cmidrule(lr){6-9}\cmidrule(l){10-13}
& Delim. & IP & \textit{Avg.} & DCSR
& Delim. & IP & \textit{Avg.} & DCSR
& Delim. & IP & \textit{Avg.} & DCSR \\
\midrule
\multicolumn{13}{@{}l}{\textit{Agent Security Bench (ASB)}} \\
\midrule
Combined Attack \citep{liu2024formalizing} & 51.0 & 49.0 & 50.0 & 36.5 & 30.5 & 26.0 & 28.3 & 16.5 & 45.5 & 48.5 & 47.0 & 37.0 \\
AutoHijacker \citep{liu2025autohijacker} & 55.0 & 55.0 & 55.0 & 41.5 & 38.5 & 39.0 & 38.8 & 25.0 & 49.5 & 49.0 & 49.2 & 40.5 \\
AgentVigil \citep{wang2025agentvigil} & 68.0 & 63.0 & 65.5 & 51.0 & 51.0 & 41.5 & 46.3 & 32.5 & 49.5 & \SecondResult{52.5} & 51.0 & 42.5 \\
A-Mem \citep{xu2025amem} & \SecondResult{73.5} & \SecondResult{63.5} & \SecondResult{68.5} & \SecondResult{55.0} & 43.0 & 37.5 & 40.3 & 27.0 & 53.0 & 51.0 & 52.0 & \SecondResult{44.5} \\
RedAgent \citep{xu2026redagent} & 70.5 & 59.5 & 65.0 & 52.5 & 48.0 & 35.5 & 41.8 & 27.0 & 54.5 & 52.0 & \SecondResult{53.3} & 44.0 \\
MARS \citep{hou2026mars} & 71.0 & 61.5 & 66.3 & 51.5 & \SecondResult{56.0} & \SecondResult{48.5} & \SecondResult{52.3} & \SecondResult{35.0} & \SecondResult{56.0} & 48.5 & 52.3 & 43.5 \\
IterInject \citep{chen2026iterinject} & 63.5 & 53.5 & 58.5 & 44.0 & 46.0 & 33.0 & 39.5 & 22.0 & 49.5 & 48.5 & 49.0 & 39.5 \\
\midrule
\textbf{\textsc{SAVOR} (Ours)} & \BestResult{84.5} & \BestResult{76.0} & \BestResult{80.3} & \BestResult{68.5} & \BestResult{59.5} & \BestResult{56.0} & \BestResult{57.8} & \BestResult{42.0} & \BestResult{57.0} & \BestResult{56.0} & \BestResult{56.5} & \BestResult{48.5} \\
\midrule
\multicolumn{13}{@{}l}{\textit{OpenClaw-IPI}} \\
\midrule
Combined Attack \citep{liu2024formalizing}
& 46.7 & 36.7 & 41.7 & 21.7
& 26.7 & 26.7 & 26.7 & 13.3
& \SecondResultO{40.0} & 36.7 & 38.3 & \SecondResultO{31.7} \\
AutoHijacker \citep{liu2025autohijacker}
& 56.7 & 61.7 & 59.2 & 40.0
& \SecondResultO{60.0} & \SecondResultO{46.7} & \SecondResultO{53.3} & \SecondResultO{43.3}
& 33.3 & 33.3 & 33.3 & 23.3 \\
AgentVigil \citep{wang2025agentvigil}
& \SecondResultO{60.0} & 53.3 & 56.7 & 31.7
& 31.7 & 20.0 & 25.8 & 11.7
& 35.0 & 40.0 & 37.5 & 28.3 \\
A-Mem \citep{xu2025amem}
& 51.7 & 41.7 & 46.7 & 23.3
& 51.7 & 38.3 & 45.0 & 28.3
& \BestResultO{53.3} & 40.0 & \SecondResultO{46.7} & 30.0 \\
RedAgent \citep{xu2026redagent}
& 48.3 & 43.3 & 45.8 & 30.0
& 50.0 & 45.0 & 47.5 & 28.3
& 26.7 & 16.7 & 21.7 & 11.7 \\
MARS \citep{hou2026mars}
& \BestResultO{81.7} & \SecondResultO{68.3} & \SecondResultO{75.0} & \SecondResultO{55.0}
& 55.0 & 43.3 & 49.2 & 30.0
& 35.0 & 18.3 & 26.7 & 10.0 \\
IterInject \citep{chen2026iterinject}
& \SecondResultO{60.0} & 61.7 & 60.8 & 40.0
& 41.7 & 26.7 & 34.2 & 16.7
& \SecondResultO{40.0} & \SecondResultO{43.3} & 41.7 & 26.7 \\
\midrule
\textbf{\textsc{SAVOR} (Ours)}
& \BestResultO{81.7} & \BestResultO{83.3} & \BestResultO{82.5} & \BestResultO{68.3}
& \BestResultO{65.0} & \BestResultO{56.7} & \BestResultO{60.8} & \BestResultO{45.0}
& \BestResultO{53.3} & \BestResultO{45.0} & \BestResultO{49.2} & \BestResultO{33.3} \\
\bottomrule
\end{tabular}
\caption{Iteration-1 Test ASR and DCSR (\%) under Delimiter (Delim.) and
Instructional Prevention (IP); \textit{Avg.} is their mean. Bold and underline:
best and second-best per block.}
\label{tab:main-results}
\end{table*}

\begin{figure*}[!t]
\centering
\includegraphics[width=0.9\textwidth]{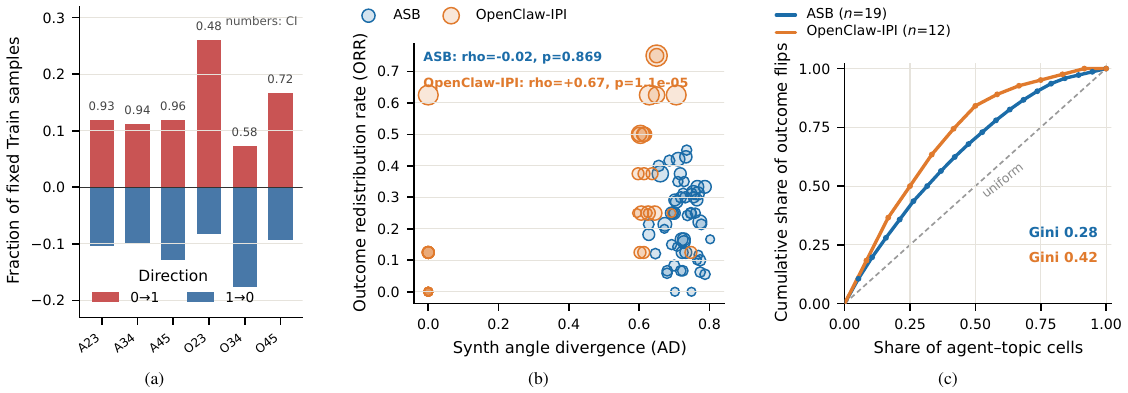}
\caption{Train-only strategy dynamics (DeepSeek-V4-Flash, Delimiter) across
\(D_k\!\rightarrow D_{k+1}\). (a) outcome flows labelled by cancellation index
(CI); (b) angle divergence (AD) against outcome redistribution rate (ORR),
marker area for absolute ASR change; (c) cumulative share of outcome flips
against agent--topic cells ranked by flips, uniform diagonal for reference.}
\label{fig:strategy-dynamics}
\end{figure*}

Table~\ref{tab:main-results} reports the iteration-1 Test results on both
benchmarks.
\textsc{SAVOR} attains the highest average ASR in all six
benchmark--target-model settings, leading the strongest baseline in each by
2.5 to 11.8 points. All entries are single-run point estimates on splits of
fixed size, so the larger margins are the ones that carry weight and the
narrowest should be read as a tie rather than an ordering. Three patterns organize the table rather than any single
entry. First, absolute difficulty is set by the target model, not by the
attack: average ASR on OpenClaw-IPI falls from 82.5\% to 49.2\% as the victim
changes from DeepSeek-V4-Flash to GPT-5.4-mini, yet \textsc{SAVOR} remains
first in every column. Second, its margin over the strongest baseline is
larger under Instructional Prevention than under Delimiter in all six
settings, indicating that learned strategies matter most where the defense is
stronger and fixed payloads degrade fastest. Third, the margin is widest
against DeepSeek-V4-Flash and narrowest against GPT-5.4-mini; because
DeepSeek-V4-Flash is also the payload generator, generator and victim are not
varied independently here, and we report this ordering as an observation
rather than a same-family transfer effect. \textsc{SAVOR} also achieves the
highest DCSR in all six settings, so the advantage survives the stricter
requirement that a single instance succeed under both defenses. These gains are not bought
with compute: At test time, \textsc{SAVOR} has the lowest attack-generation cost,
below \(10^{-4}\)~USD per sample on both benchmarks
(Appendix~E of the supplementary material).


\paragraph{Train ASR trajectory.}
Aggregate Train ASR rises over the five offline iterations on both benchmarks,
but iteration~1 and iteration~2 draw on different pools, so only the
iteration~2--5 segments are comparable; the full trajectory is plotted in
Appendix~F of the supplementary material.

\paragraph{ASB: buffered regime.}
Figure~\ref{fig:strategy-dynamics} aligns each round's Analyzer, Synthesizer,
and Memory artifacts with the next fixed-pool Train transition. The unit of
analysis is the cell transition. The outcome redistribution rate (ORR) is the
fraction of matched samples whose outcome changes across a transition; the
cancellation index (CI) is one minus the normalized net imbalance between the
failure-to-success and success-to-failure flows; and angle divergence (AD) is
one minus the mean symmetric similarity of the ATK, DEF, and JNT strategy
texts. These Train-only observations come from one victim model under one
defense, and the associations reported below are exploratory rather than
planned tests. Across
iterations~2--5, ASB ASR changes from 74.1\% to 75.6\%, 76.9\%, and 75.9\%.
However, 21.2--24.7\% of matched samples change outcome in each transition,
while cancellation indices of 0.930--0.962 show that failure-to-success and
success-to-failure flows nearly offset. Angle divergence is not associated
with downstream outcome redistribution (\(\rho=-0.023\), \(p=0.869\)), and
memory novelty is similarly weakly associated (\(\rho=0.155\), \(p=0.248\)).
In these transitions ASB therefore shows a buffered pattern, in which
continued strategy reconfiguration produces substantial sample-level movement
but little net ASR change.
\begin{table*}[!t]
\centering
\renewcommand{\arraystretch}{0.9}
\small
\begin{tabular}{@{}p{0.47\textwidth}@{\hspace{0.02\textwidth}}p{0.47\textwidth}@{}}
\begin{tabular*}{\linewidth}[t]{@{\extracolsep{\fill}}>{\hlstrut}lcc@{}}
\toprule
\textit{(a)} Variant & ASB & OpenClaw-IPI \\
\midrule
w/o Memory Refiner & 80.5 ($-4.0$) & 78.3 ($-3.4$) \\
Full \textsc{SAVOR} & \BestResult{84.5} & \BestResultO{81.7} \\
\bottomrule
\end{tabular*}
&
\begin{tabular*}{\linewidth}[t]{@{\extracolsep{\fill}}>{\hlstrut}lccc@{}}
\toprule
\textit{(b)} Benchmark & Best base. & Delim.$\to$IP & IP opt. \\
\midrule
ASB & 63.5 & 72.5 & \BestResult{76.0} \\
OpenClaw-IPI & 68.3 & 68.3 & \BestResultO{83.3} \\
\bottomrule
\end{tabular*}
\\[-4pt]
\begin{tabular*}{\linewidth}[t]{@{\extracolsep{\fill}}>{\hlstrut}lcc@{}}
\toprule
\textit{(c)} Attack Generator & ASB & OpenClaw-IPI \\
\midrule
DeepSeek-V4-Flash & \BestResult{84.5} & 81.7 \\
Qwen3.5-Flash Thinking & 79.5 & 81.7 \\
GPT-5.4-mini & 63.0 & \BestResultO{88.3} \\
\bottomrule
\end{tabular*}
&
\begin{tabular*}{\linewidth}[t]{@{\extracolsep{\fill}}>{\hlstrut}lcc@{}}
\toprule
\textit{(d)} Synthesis angle & ASB & OpenClaw-IPI \\
\midrule
ATK only & 82.5 ($-2.0$) & 66.7 ($-15.0$) \\
DEF only & 83.5 ($-1.0$) & 66.7 ($-15.0$) \\
JNT only & 79.0 ($-5.5$) & 81.7 ($\pm0.0$) \\
\bottomrule
\end{tabular*}
\end{tabular}
\caption{Iteration-1 analyses (DeepSeek-V4-Flash, Delimiter): (a) refiner
ablation; (b) cross-defense transfer; (c) Attack Generator choice; (d) fixed
synthesis angle, which also isolates the success-only (ATK) and failure-only
(DEF) reflection sources. Parentheses: points relative to Full \textsc{SAVOR}.}
\label{tab:analysis-results}
\end{table*}

\begin{table}[!t]
\centering
\renewcommand{\arraystretch}{0.96}
\begin{tabular*}{0.86\linewidth}{@{\extracolsep{\fill}}lcc@{}}
\toprule
Refiner operation & ASB & OpenClaw-IPI \\
\midrule
Revise    & \textbf{226} (61.6) & 3 (1.8) \\
Keep      & 44 (12.0)  & 57 (34.5) \\
Add       & 32 (8.7)   & \textbf{63} (38.2) \\
Retire    & 61 (16.6)  & 42 (25.5) \\
Condition & 4 (1.1)    & 0 (0.0) \\
\midrule
Total operations  & 367 & 165 \\
Revisions / edits & \textbf{70.0} & 2.8 \\
Turnover / edits  & 28.8 & \textbf{97.2} \\
\bottomrule
\end{tabular*}
\caption{Memory Refiner edit composition over the three fixed-pool Train
transitions (DeepSeek-V4-Flash, Delimiter); counts (column \%).
\textit{Edits} excludes Keep; turnover is Add\,+\,Retire.}
\label{tab:memory-edits}
\end{table}
\paragraph{OpenClaw-IPI: coupled regime.}
OpenClaw-IPI shows a more selectively coupled pattern
(Figure~\ref{fig:strategy-dynamics}).
Its fixed-pool ORR remains 25.0--34.4\%, but these transitions retain more net
direction than in ASB (CI \(=0.485\)--0.720). More importantly, Synth angle
divergence is positively associated with downstream redistribution
(\(\rho=0.668\), \(p=1.15\times10^{-5}\)). Analyzer evidence dispersion is
also positively associated with redistribution (\(\rho=0.449\), \(p=0.006\)),
whereas stronger Synth grounding is negatively associated
(\(\rho=-0.465\), \(p=0.004\)). These associations do not establish causality,
but they show that OpenClaw-IPI's ASR fluctuations coincide with structured
changes in the strategy representation rather than undirected round-to-round
noise. These aggregate flows are not diffuse: on both benchmarks they
are unevenly distributed over agent--topic cells, and more so on
OpenClaw-IPI: the busiest quarter of cells carries 50\% of all outcome flips
there against 44\% on ASB, and the busiest half carries 84\% against 73\%
(Gini 0.42 against 0.28). Appendix~F localizes the flows cell by cell.

\paragraph{Ablation and defense transfer.}
Table~\ref{tab:analysis-results}(a) isolates the Memory Refiner after one
round. Removing it costs 4.0 points on ASB and 3.4 on OpenClaw-IPI, so
consolidation contributes but is not the main source of the gain; this test is
limited to DeepSeek-V4-Flash under Delimiter. Table~\ref{tab:analysis-results}(b)
applies memory learned under Delimiter to Instructional Prevention. Transfer
beats the strongest IP baseline on both benchmarks while recovering 95\% of
the defense-specific optimum on ASB but only 82\% on OpenClaw-IPI: strategies
carry across defenses, and they lose more in the executable setting than in
the simulated one. Reverse transfer is untested.
Table~\ref{tab:analysis-results}(d) replaces validation-guided angle selection
with a single fixed angle, which also isolates the two reflection sources:
ATK uses only success reflections, whereas DEF uses only failure reflections. Its value is avoiding a poor angle rather than
exceeding the best one: committing to ATK or DEF alone costs 15.0 points on
OpenClaw-IPI, whereas fixed JNT ties selection there while losing 5.5 points on
ASB. No fixed angle is best on both benchmarks, and selection operates per
cell rather than globally, so it recovers the better angle without knowing
which one it is in advance.

\paragraph{Memory edit composition.}
Table~\ref{tab:memory-edits} decomposes what the Refiner actually does across
the three fixed-pool transitions. The two benchmarks sit at opposite ends:
70.0\% of ASB edits revise an existing strategy, against 2.8\% on
OpenClaw-IPI, where 97.2\% of edits instead add or retire. Memory in the
simulated setting therefore consolidates, whereas memory in the executable
setting turns over almost completely each round --- a second signature of the
same buffered versus selectively coupled contrast seen in the outcome flows.

\paragraph{Attack Generator choice.}
Table~\ref{tab:analysis-results}(c) reuses the selected strategies while
changing the Attack Generator. No generator dominates: GPT-5.4-mini is the
weakest on ASB and the strongest on OpenClaw-IPI (63.0\% against 88.3\%),
while Qwen3.5-Flash Thinking falls between the two on both. Generator choice
therefore interacts with the benchmark instead of producing a single ordering,
so the strategies are not tied to one generator but the achievable ceiling
is.

\paragraph{Benchmark scope and outlook.}
OpenClaw-IPI holds out payload families, so its Test split measures transfer to
unseen attack \emph{goals}; holding out tool families as well would extend the
same execution-grounded protocol to unseen capability surfaces. Its compact
inventory likewise supports point estimates, and enlarging it would make
repeated-run variance estimable. Both are natural extensions of the benchmark
rather than obstacles to the present comparison, and we see them, together with
reverse defense transfer and further victim-model families, as the most
informative next experiments.

\section{Conclusion}

We introduced \textsc{SAVOR}, a pre-deployment adaptation framework that
turns successful and failed training trajectories into context-conditioned
attack strategies, selects effective angles through validation, and
consolidates them into a reusable memory for strict one-query deployment
against unseen attacker tools. \textsc{SAVOR} attains the highest average ASR in
all six benchmark--target-model settings, its memory transfers across
defenses, and its Train dynamics separate a largely offsetting pattern on ASB
from a more directional one on OpenClaw-IPI. Attack knowledge distilled before
deployment therefore transfers to attacker tools held out on ASB and to attack
goals held out on OpenClaw-IPI, and defenses should be evaluated against such
attackers, not only against fixed payloads or online adaptation.

{\small
\bibliography{aaai2027}

\begin{thebibliography}{34}
\providecommand{\natexlab}[1]{#1}

\bibitem[{Bhagwatkar et~al.(2025)Bhagwatkar, Kasa, Puri, Huang, Rish, Taylor,
  Dvijotham, and Lacoste}]{bhagwatkar2025firewalls}
Bhagwatkar, R.; Kasa, K.; Puri, A.; Huang, G.; Rish, I.; Taylor, G.~W.;
  Dvijotham, K.~D.; and Lacoste, A. 2025.
\newblock Indirect Prompt Injections: Are Firewalls All You Need, or Stronger
  Benchmarks?
\newblock arXiv:2510.05244.

\bibitem[{Chao et~al.(2025)Chao, Robey, Dobriban, Hassani, Pappas, and
  Wong}]{chao2025pair}
Chao, P.; Robey, A.; Dobriban, E.; Hassani, H.; Pappas, G.~J.; and Wong, E.
  2025.
\newblock Jailbreaking Black Box Large Language Models in Twenty Queries.
\newblock In \emph{2025 IEEE Conference on Secure and Trustworthy Machine
  Learning (SaTML)}, 23--42. IEEE.

\bibitem[{Chen et~al.(2026)Chen, Chen, Luo, Xu, Huang, Sun, and
  Jiang}]{chen2026iterinject}
Chen, Z.; Chen, J.; Luo, L.; Xu, K.; Huang, X.; Sun, T.; and Jiang, X. 2026.
\newblock {IterInject}: Indirect Prompt Injection Against {LLM} Agents via
  Feedback-Guided Iterative Optimization.
\newblock arXiv:2605.24659.

\bibitem[{Debenedetti et~al.(2024)Debenedetti, Zhang, Balunovic,
  Beurer-Kellner, Fischer, and Tram{\`e}r}]{debenedetti2024agentdojo}
Debenedetti, E.; Zhang, J.; Balunovic, M.; Beurer-Kellner, L.; Fischer, M.; and
  Tram{\`e}r, F. 2024.
\newblock {AgentDojo}: A Dynamic Environment to Evaluate Prompt Injection
  Attacks and Defenses for {LLM} Agents.
\newblock In \emph{Advances in Neural Information Processing Systems},
  volume~37, 82895--82920. Curran Associates, Inc.

\bibitem[{Dziemian et~al.(2026)Dziemian, Lin, Fu, Nowak, Winter, Jones, Zou,
  Ahmad, Chaudhuri, Chennabasappa, Davies, Deason, Edelman, Emek, Evtimov,
  Gust, Hamin, He, Krawiecka, Patana, Perry, Peterson, Qi, Rando, Wang, Wang,
  Whitman, Winsor, Zharmagambetov, Fredrikson, and Kolter}]{do2026competition}
Dziemian, M.; Lin, M.; Fu, X.; Nowak, M.; Winter, N.; Jones, E.; Zou, A.;
  Ahmad, L.; Chaudhuri, K.; Chennabasappa, S.; Davies, X.; Deason, L.; Edelman,
  B.~L.; Emek, T.; Evtimov, I.; Gust, J.; Hamin, M.; He, K.; Krawiecka, K.;
  Patana, R.; Perry, N.; Peterson, T.; Qi, X.; Rando, J.; Wang, Z.; Wang, Z.;
  Whitman, S.; Winsor, E.; Zharmagambetov, A.; Fredrikson, M.; and Kolter, Z.
  2026.
\newblock How Vulnerable Are {AI} Agents to Indirect Prompt Injections?
  Insights from a Large-Scale Public Competition.
\newblock arXiv:2603.15714.

\bibitem[{Flavell(1979)}]{flavell1979metacognition}
Flavell, J.~H. 1979.
\newblock Metacognition and Cognitive Monitoring: A New Area of
  Cognitive-Developmental Inquiry.
\newblock \emph{American Psychologist}, 34(10): 906--911.

\bibitem[{Greshake et~al.(2023)Greshake, Abdelnabi, Mishra, Endres, Holz, and
  Fritz}]{greshake2023indirect}
Greshake, K.; Abdelnabi, S.; Mishra, S.; Endres, C.; Holz, T.; and Fritz, M.
  2023.
\newblock Not What You've Signed Up For: Compromising Real-World
  {LLM}-Integrated Applications with Indirect Prompt Injection.
\newblock In \emph{Proceedings of the 16th ACM Workshop on Artificial
  Intelligence and Security}, 79--90.

\bibitem[{Hou et~al.(2026)Hou, Qu, Wang, Gong, Guo, and Liu}]{hou2026mars}
Hou, X.; Qu, B.; Wang, W.; Gong, P.; Guo, Q.; and Liu, Y. 2026.
\newblock Learn Like Humans: Use Meta-cognitive Reflection for Efficient
  Self-Improvement.
\newblock In \emph{Proceedings of the 64th Annual Meeting of the Association
  for Computational Linguistics (Volume 1: Long Papers)}, 28802--28824. San
  Diego, California, United States: Association for Computational Linguistics.

\bibitem[{Kapur(2014)}]{kapur2014productive}
Kapur, M. 2014.
\newblock Productive Failure in Learning Math.
\newblock \emph{Cognitive Science}, 38(5): 1008--1022.

\bibitem[{Liu et~al.(2025)Liu, Jha, McDaniel, Li, and
  Xiao}]{liu2025autohijacker}
Liu, X.; Jha, S.; McDaniel, P.; Li, B.; and Xiao, C. 2025.
\newblock {AutoHijacker}: Automatic Indirect Prompt Injection Against Black-box
  {LLM} Agents.
\newblock OpenReview preprint.

\bibitem[{Liu et~al.(2023)Liu, Deng, Li, Wang, Wang, Wang, Zhang, Liu, Wang,
  Zheng, Zhang, and Liu}]{liu2023prompt}
Liu, Y.; Deng, G.; Li, Y.; Wang, K.; Wang, Z.; Wang, X.; Zhang, T.; Liu, Y.;
  Wang, H.; Zheng, Y.; Zhang, L.~Y.; and Liu, Y. 2023.
\newblock Prompt Injection attack against {LLM}-integrated Applications.
\newblock arXiv:2306.05499.

\bibitem[{Liu et~al.(2024)Liu, Jia, Geng, Jia, and Gong}]{liu2024formalizing}
Liu, Y.; Jia, Y.; Geng, R.; Jia, J.; and Gong, N.~Z. 2024.
\newblock Formalizing and Benchmarking Prompt Injection Attacks and Defenses.
\newblock In \emph{33rd USENIX Security Symposium (USENIX Security 24)},
  1831--1847. Philadelphia, PA: USENIX Association.

\bibitem[{Ma et~al.(2026)Ma, Li, Xiao, Yu, Zhang, and
  Vorobeychik}]{ma2026autodojo}
Ma, X.; Li, T.; Xiao, C.; Yu, Z.; Zhang, N.; and Vorobeychik, Y. 2026.
\newblock {AutoDojo}: Adaptive Black-Box Attacks Reveal the Limits of {IPI}
  Defenses and Task-Specification Effects in {LLM} Agents.
\newblock arXiv:2606.15057.

\bibitem[{Madaan et~al.(2023)Madaan, Tandon, Gupta, Hallinan, Gao, Wiegreffe,
  Alon, Dziri, Prabhumoye, Yang, Gupta, Majumder, Hermann, Welleck,
  Yazdanbakhsh, and Clark}]{madaan2023selfrefine}
Madaan, A.; Tandon, N.; Gupta, P.; Hallinan, S.; Gao, L.; Wiegreffe, S.; Alon,
  U.; Dziri, N.; Prabhumoye, S.; Yang, Y.; Gupta, S.; Majumder, B.~P.; Hermann,
  K.; Welleck, S.; Yazdanbakhsh, A.; and Clark, P. 2023.
\newblock {Self-Refine}: Iterative Refinement with Self-Feedback.
\newblock In \emph{Advances in Neural Information Processing Systems},
  volume~36, 46534--46594. Curran Associates, Inc.

\bibitem[{Mehrotra et~al.(2024)Mehrotra, Zampetakis, Kassianik, Nelson,
  Anderson, Singer, and Karbasi}]{mehrotra2024tree}
Mehrotra, A.; Zampetakis, M.; Kassianik, P.; Nelson, B.; Anderson, H.; Singer,
  Y.; and Karbasi, A. 2024.
\newblock Tree of Attacks: Jailbreaking Black-Box {LLM}s Automatically.
\newblock In \emph{Advances in Neural Information Processing Systems},
  volume~37, 61065--61105. Curran Associates, Inc.

\bibitem[{{Qwen Team}(2026)}]{qwenteam2026qwen36}
{Qwen Team}. 2026.
\newblock {Qwen3.6-27B}: Flagship-Level Coding in a {27B} Dense Model.
\newblock Qwen technical blog.
\newblock Accessed: 2026-07-29.

\bibitem[{Shayoni et~al.(2026)Shayoni, Shoaib, Hossain, and
  Mridha}]{shayoni2026netinjectbench}
Shayoni, R.~K.; Shoaib, M.~F.; Hossain, S. M.~A.; and Mridha, M.~F. 2026.
\newblock {NetInjectBench}: Benchmarking Indirect Prompt Injection in
  Tool-Using Large Language Model Agents for Network Operations.
\newblock arXiv:2607.10490.

\bibitem[{Shinn et~al.(2023)Shinn, Cassano, Gopinath, Narasimhan, and
  Yao}]{shinn2023reflexion}
Shinn, N.; Cassano, F.; Gopinath, A.; Narasimhan, K.; and Yao, S. 2023.
\newblock Reflexion: Language Agents with Verbal Reinforcement Learning.
\newblock In \emph{Advances in Neural Information Processing Systems},
  volume~36, 8634--8652. Curran Associates, Inc.

\bibitem[{Syros et~al.(2026)Syros, Rose, Grinstead, Kerschbaumer, Robertson,
  Nita-Rotaru, and Oprea}]{syros2026muzzle}
Syros, G.; Rose, E.; Grinstead, B.; Kerschbaumer, C.; Robertson, W.;
  Nita-Rotaru, C.; and Oprea, A. 2026.
\newblock {MUZZLE}: Adaptive Agentic Red-Teaming of Web Agents Against Indirect
  Prompt Injection Attacks.
\newblock In \emph{35th USENIX Security Symposium (USENIX Security 26)}.
  Baltimore, MD: USENIX Association.

\bibitem[{Wang et~al.(2026)Wang, Zhang, Zhang, Wang, Wang, Gao, Wei, Chen, and
  Lim}]{wang2026adaptools}
Wang, C.; Zhang, J.; Zhang, Z.; Wang, Z.; Wang, Y.; Gao, J.; Wei, T.; Chen, Z.;
  and Lim, W. Y.~B. 2026.
\newblock {AdapTools}: Adaptive Tool-based Indirect Prompt Injection Attacks on
  Agentic {LLM}s.
\newblock arXiv:2602.20720.

\bibitem[{Wang et~al.(2024)Wang, Ma, Feng, Zhang, Yang, Zhang, Chen, Tang,
  Chen, Lin, Zhao, Wei, and Wen}]{wang2024survey}
Wang, L.; Ma, C.; Feng, X.; Zhang, Z.; Yang, H.; Zhang, J.; Chen, Z.; Tang, J.;
  Chen, X.; Lin, Y.; Zhao, W.~X.; Wei, Z.; and Wen, J.-R. 2024.
\newblock A Survey on Large Language Model based Autonomous Agents.
\newblock \emph{Frontiers of Computer Science}, 18(6): 186345.

\bibitem[{Wang et~al.(2025)Wang, Siu, Ye, Shi, Nie, Zhao, Wang, Guo, and
  Song}]{wang2025agentvigil}
Wang, Z.; Siu, V.; Ye, Z.; Shi, T.; Nie, Y.; Zhao, X.; Wang, C.; Guo, W.; and
  Song, D. 2025.
\newblock {AGENTVIGIL}: Automatic Black-Box Red-teaming for Indirect Prompt
  Injection against {LLM} Agents.
\newblock In \emph{Findings of the Association for Computational Linguistics:
  EMNLP 2025}, 23159--23172. Suzhou, China: Association for Computational
  Linguistics.

\bibitem[{Xiang et~al.(2026)Xiang, Zagieboylo, Ghosh, Kariyappa, Greshake,
  Xiao, Xiao, and Suh}]{xiang2026architecting}
Xiang, C.; Zagieboylo, D.; Ghosh, S.; Kariyappa, S.; Greshake, K.; Xiao, H.;
  Xiao, C.; and Suh, G.~E. 2026.
\newblock Architecting Secure {AI} Agents: Perspectives on System-Level
  Defenses Against Indirect Prompt Injection Attacks.
\newblock arXiv:2603.30016.

\bibitem[{Xie et~al.(2024)Xie, Zhang, Chen, Li, Zhao, Cao, Hua, Cheng, Shin,
  Lei, Liu, Xu, Zhou, Savarese, Xiong, Zhong, and Yu}]{xie2024osworld}
Xie, T.; Zhang, D.; Chen, J.; Li, X.; Zhao, S.; Cao, R.; Hua, T.~J.; Cheng, Z.;
  Shin, D.; Lei, F.; Liu, Y.; Xu, Y.; Zhou, S.; Savarese, S.; Xiong, C.; Zhong,
  V.; and Yu, T. 2024.
\newblock {OSWorld}: Benchmarking Multimodal Agents for Open-Ended Tasks in
  Real Computer Environments.
\newblock In \emph{Advances in Neural Information Processing Systems},
  volume~37, 52040--52094. Curran Associates, Inc.

\bibitem[{Xu et~al.(2026)Xu, Zhang, Wang, Xiao, Zheng, Ba, and
  Ren}]{xu2026redagent}
Xu, H.; Zhang, W.; Wang, Z.; Xiao, F.; Zheng, R.; Ba, Z.; and Ren, K. 2026.
\newblock {RedAgent}: An Autonomous Agent for Context-Aware Red Teaming of
  {LLM} Jailbreaks.
\newblock \emph{IEEE Transactions on Dependable and Secure Computing}, 23(3):
  6506--6521.

\bibitem[{Xu et~al.(2025)Xu, Liang, Mei, Gao, Tan, and Zhang}]{xu2025amem}
Xu, W.; Liang, Z.; Mei, K.; Gao, H.; Tan, J.; and Zhang, Y. 2025.
\newblock {A-Mem}: Agentic Memory for {LLM} Agents.
\newblock In \emph{Advances in Neural Information Processing Systems},
  volume~38, 17577--17604. Curran Associates, Inc.

\bibitem[{Yang et~al.(2024)Yang, Jimenez, Wettig, Lieret, Yao, Narasimhan, and
  Press}]{yang2024sweagent}
Yang, J.; Jimenez, C.~E.; Wettig, A.; Lieret, K.; Yao, S.; Narasimhan, K.; and
  Press, O. 2024.
\newblock {SWE-agent}: Agent-Computer Interfaces Enable Automated Software
  Engineering.
\newblock In \emph{Advances in Neural Information Processing Systems},
  volume~37, 50528--50652. Curran Associates, Inc.

\bibitem[{Yao et~al.(2023)Yao, Zhao, Yu, Du, Shafran, Narasimhan, and
  Cao}]{yao2023react}
Yao, S.; Zhao, J.; Yu, D.; Du, N.; Shafran, I.; Narasimhan, K.~R.; and Cao, Y.
  2023.
\newblock ReAct: Synergizing Reasoning and Acting in Language Models.
\newblock In \emph{The Eleventh International Conference on Learning
  Representations}.

\bibitem[{Yi et~al.(2025)Yi, Xie, Zhu, Kiciman, Sun, Xie, and Wu}]{yi2025bipia}
Yi, J.; Xie, Y.; Zhu, B.; Kiciman, E.; Sun, G.; Xie, X.; and Wu, F. 2025.
\newblock Benchmarking and Defending Against Indirect Prompt Injection Attacks
  on Large Language Models.
\newblock In \emph{Proceedings of the 31st ACM SIGKDD Conference on Knowledge
  Discovery and Data Mining}, 1809--1820. Association for Computing Machinery.

\bibitem[{Zhan et~al.(2025)Zhan, Fang, Panchal, and Kang}]{zhan2025adaptive}
Zhan, Q.; Fang, R.; Panchal, H.~S.; and Kang, D. 2025.
\newblock Adaptive Attacks Break Defenses Against Indirect Prompt Injection
  Attacks on {LLM} Agents.
\newblock In \emph{Findings of the Association for Computational Linguistics:
  NAACL 2025}, 7116--7132. Albuquerque, New Mexico: Association for
  Computational Linguistics.

\bibitem[{Zhan et~al.(2024)Zhan, Liang, Ying, and Kang}]{zhan2024injecagent}
Zhan, Q.; Liang, Z.; Ying, Z.; and Kang, D. 2024.
\newblock {InjecAgent}: Benchmarking Indirect Prompt Injections in
  Tool-Integrated Large Language Model Agents.
\newblock In \emph{Findings of the Association for Computational Linguistics:
  ACL 2024}, 10471--10506. Bangkok, Thailand: Association for Computational
  Linguistics.

\bibitem[{Zhang et~al.(2025)Zhang, Huang, Mei, Yao, Wang, Zhan, Wang, and
  Zhang}]{zhang2025asb}
Zhang, H.; Huang, J.; Mei, K.; Yao, Y.; Wang, Z.; Zhan, C.; Wang, H.; and
  Zhang, Y. 2025.
\newblock Agent Security Bench ({ASB}): Formalizing and Benchmarking Attacks
  and Defenses in {LLM}-based Agents.
\newblock In \emph{The Thirteenth International Conference on Learning
  Representations}.

\bibitem[{Zhang et~al.(2026)Zhang, Xu, Wang, Guo, Xu, Xiao, Guan, Fan, Liu,
  Liu, and Hu}]{zhang2026agentsentry}
Zhang, T.; Xu, Y.; Wang, J.; Guo, K.; Xu, X.; Xiao, B.; Guan, Q.; Fan, J.; Liu,
  J.; Liu, Z.; and Hu, H. 2026.
\newblock {AgentSentry}: Mitigating Indirect Prompt Injection in {LLM} Agents
  via Temporal Causal Diagnostics and Context Purification.
\newblock arXiv:2602.22724.

\bibitem[{Zhou et~al.(2024)Zhou, Xu, Zhu, Zhou, Lo, Sridhar, Cheng, Ou, Bisk,
  Fried, Alon, and Neubig}]{zhou2024webarena}
Zhou, S.; Xu, F.~F.; Zhu, H.; Zhou, X.; Lo, R.; Sridhar, A.; Cheng, X.; Ou, T.;
  Bisk, Y.; Fried, D.; Alon, U.; and Neubig, G. 2024.
\newblock WebArena: A Realistic Web Environment for Building Autonomous Agents.
\newblock In \emph{The Twelfth International Conference on Learning
  Representations}.

\end{thebibliography}
}


\appendix
\setcounter{secnumdepth}{1}


\section{Appendix Scope and Evidence Status}
\label{app:scope}

\subsection{Reading Guide and Terminology}

This appendix separates the experimental protocol from the empirical evidence.
It first specifies the information-access contract, algorithm, data
construction, evaluation rules, baseline adaptations, and resource accounting.
It then reports disaggregated results only for conditions supported by
traceable experimental artifacts.  Analyses without completed evaluations are
identified explicitly; no values are estimated and no empirical conclusions
are drawn from them.

We use \emph{round} for one offline \textsc{SAVOR} learning cycle and
\emph{iteration-$k$ snapshot} for the state after round~$k$.  The
main paper reports the \emph{iteration-1 snapshot}; the four later
snapshots are used only to characterize trends and diagnostic trajectories.  A
\emph{victim query} is one victim-agent execution.  Accordingly,
\emph{one-query Test} means that each Test sample receives exactly one such
execution; it does not imply one attacker-side model call or equal total
offline optimization cost.  We use \emph{strategy memory} for the state updated
across offline rounds and \emph{frozen strategy library} for the strategy
content exposed to generation after offline learning ends.

\subsection{Evidence Status}

Table~\ref{tab:appendix-evidence-status} distinguishes completed experimental
conditions from analyses that require additional experiments.  Each completed
ASB and OpenClaw-IPI condition is a single run, supporting a point estimate but
not statistical significance or stability across independent runs.

\begin{table*}[!t]
\centering

\footnotesize
\setlength{\tabcolsep}{5pt}
\begin{tabular}{llll}
\toprule
Benchmark or analysis & Victim / setting & Defense or scope & Status \\
\midrule
ASB main comparison & DeepSeek-V4-Flash & Delimiter & Completed (single run) \\
ASB main comparison & DeepSeek-V4-Flash & Instructional Prevention & Completed (single run) \\
ASB main comparison & Additional victim models & Both prompt defenses & Completed (single run) \\
OpenClaw-IPI comparison & All reported victim models & Both prompt defenses & Completed (single run) \\
Memory Refiner ablation & DeepSeek-V4-Flash & Delimiter & Completed (single run) \\
Memory Refiner ablation & DeepSeek-V4-Flash & Instructional Prevention & Completed (single run) \\
Repeated-run uncertainty & All conditions & Multiple independent runs & Further experiments required \\
Validation reuse audit & \textsc{SAVOR} & Untouched selection-audit split & Further experiments required \\
Budget-matched ablations & \textsc{SAVOR} and baselines & Matched total optimization budget & Further experiments required \\
\bottomrule
\end{tabular}
\caption{Completion status of the experiments and analyses considered in this
study.  ``Further experiments required'' denotes an analysis that cannot be
recovered from the completed artifacts.}
\label{tab:appendix-evidence-status}
\end{table*}


\section{Formal Threat Model and Information Boundaries}
\label{app:threat-model}

\subsection{Editable Observation Channel}

We study observation-level indirect prompt injection (IPI), referred to as
OPI in the released ASB implementation.  For a benign
user task \(q_i\), the victim agent receives a fixed system prompt, tool
repertoire \(\mathcal{T}_i\), and a sequence of observations.  The attacker may
control only the injected text delivered through the untrusted observation channel;
it cannot modify the system prompt, user task, tool schemas, defense wrapper,
agent memory, or runtime.  Success is evaluated from the victim execution
rather than from an attacker-side prediction.  This scope excludes direct
prompt injection and any attack that changes the victim implementation or
tools.

The abstract threat model controls one observation slot.  The benchmark
harnesses instantiate that slot differently: the released ASB harness replays
the same generated payload in every non-final observation, whereas OpenClaw-IPI
injects it into exactly one dynamically produced tool output.

The public target context is richer than victim-internals access.  In
Validation and Test, the payload-generation path can receive the visible user
task, public agent role, attacker-tool name and description, a generic
Combined Attack structure used as a seed reference, and the sample's internal
attack objective.  We therefore call this setting
\emph{victim-internals-black-box and target-metadata-visible}.  The internal
objective supplies the intended action to the generator, but the raw
\texttt{attack\_goal} field is removed from the public attacker-tool record
before the victim is run; the victim receives only the generated observation
payload.

\subsection{Phase-Specific Information Access}

Table~\ref{tab:phase-access} gives the information boundary for the adaptive
components of \textsc{SAVOR}.  ``Attacker-side only'' denotes information
available to attacker-side components but removed from the victim-facing tool
record before execution.  Training is white-box.  Validation and Test are
black-box with respect to victim internals, but Validation outcomes are reused
during offline development.

\begin{table*}[!t]
\centering

\footnotesize
\setlength{\tabcolsep}{4.5pt}
\begin{tabular}{p{0.36\textwidth}p{0.17\textwidth}p{0.18\textwidth}p{0.18\textwidth}}
\toprule
Information & Train & Validation & Test \\
\midrule
Visible user task and public agent role & Yes & Yes & Yes \\
Attacker-tool name and public description & Yes & Yes & Yes \\
Generic Combined Attack structure & Available to generation & Yes & Yes \\
Sample-specific attack objective & Yes & Yes, attacker-side only; removed before victim execution & Yes, attacker-side only; removed before victim execution \\
Injected payload being analyzed or generated & Yes & Yes & Yes \\
Victim trajectory, thinking, or trust text & Defender Analyzer only, using sanitized failed Train trajectories & No & No \\
Train success/failure outcome & Yes, with analyst-specific routing & No & No \\
Current Validation outcome or ASR & Not applicable & Winner selection after all candidate executions; optional conflict comparison & No \\
Current Test response, outcome, or ASR & Not applicable & Not applicable & No adaptive access; report only \\
Cross-round strategy memory & Read by later-round generation & Updated after winner selection & Frozen read-only input \\
\bottomrule
\end{tabular}
\caption{Phase-specific information available to the adaptive \textsc{SAVOR}
components.  Validation execution produces an outcome for aggregate winner
selection; it is not passed back to the payload generator for the evaluated
sample.  Test outcomes are report-only.}
\label{tab:phase-access}
\end{table*}

The Train Analyzer deliberately separates evidence views.  Its attacker view
analyzes successful injected payloads and their outcome labels without victim
thinking, trust text, or trajectories.  Its defender view analyzes sanitized
failed trajectories and excludes successful samples.  This separation is
enforced when the prompt packet is rendered, not merely requested in prose.
The current-round Synthesizer then consumes only the grouped Analyzer outputs;
it does not read previous strategy memory.  Cross-round strategy memory is
updated only after Validation through the Memory Refiner.  Separately, each
later-round Analyzer consumes a newly executed Train allocation rather than
previous strategy memory directly.

Validation is development data rather than an untouched evaluation set.  For
each agent--topic cell, the three ATK, DEF, and JNT candidates are executed
on the Validation tasks and aggregated into candidate-specific ASR.  The
winner selector writes one current-round winner record.  That record can alter
the strategy memory used by later offline rounds.  The implementation also
contains an optional conflict-comparison path in which the Memory Refiner may
request additional Validation evidence for directly conflicting strategies.
Whether this path triggered in a particular result tree must be established
from its canonical logs and included in the query ledger; its default
availability is not evidence that extra queries were executed.

\subsection{Module Read/Write Contract}

Table~\ref{tab:module-permissions} makes the state boundary explicit.  Private
debug traces may contain richer generation context, but they are stored under
run-log directories and are neither written into the public attack payload
file nor consumed by the Test-time adaptive path.

\begin{table*}[!t]
\centering

\footnotesize
\setlength{\tabcolsep}{4pt}
\begin{tabular}{p{0.19\textwidth}p{0.34\textwidth}p{0.34\textwidth}}
\toprule
Component & Reads & Writes \\
\midrule
Train Analyzer & Train payloads, tasks, roles, outcomes; sanitized failed trajectory for DEF only & ATK and DEF analysis records \\
Topic Aggregator & Current ATK/DEF records and fixed tool--topic mapping & Projected agent--topic buckets containing only Analyzer contract fields \\
Synthesizer & Current agent--topic buckets & Current ATK/DEF/JNT candidate strategies \\
Validation payload generator & Current candidate, public target metadata, internal objective, generic seed structure & One candidate payload per angle and sample; private debug trace \\
Validation evaluator & Candidate payload, victim runtime, defense, task and tools & Executed outcomes and aggregate inputs \\
Winner selector & Current candidate-specific Validation counts and ASR & Current winner and diagnostic records \\
Memory Refiner & Previous memory, current winner-angle strategies, optional conflict evidence & Updated cross-round memory and retired avoid lessons \\
Test Strategy Selector & Frozen active memory and public target metadata & One sample-specific strategy and private selection log \\
Attack Generator & Selected strategy, public target metadata, internal objective, generic seed structure & Public payload plus private generation diagnostics \\
Victim and reporter & Public payload/task/tools; executed Test events & Report artifacts only; no adaptive state \\
\bottomrule
\end{tabular}
\caption{Read/write contract for the current \textsc{SAVOR} pipeline.  ``Current''
means the current offline round; ``previous'' means cross-round state.}
\label{tab:module-permissions}
\end{table*}

\subsection{Frozen Test State and One-Query Execution}

Before official Test execution, the iteration-1 strategy memory, winner
diagnostics, prompt templates, model routing, defenses, and generation
hyperparameters are fixed.  For each held-out Test tool--task pair, the
Strategy Selector produces one strategy from the frozen memory and the Attack
Generator produces one payload.  Output-format parsing may retry the same
attacker-side prompt, but this retry does not query the victim and receives no
victim feedback.  The resulting payload is evaluated by exactly one
victim-agent execution.  The response, trajectory, executed-call outcome, and
ASR from that execution are not returned to the selector, generator,
Synthesizer, winner selector, or Memory Refiner.

Consequently, the supported efficiency statement is a deployment-time
boundary: one victim query per Test sample and no Test-time victim-feedback
cycle.  It is not a claim of one total model call, equal offline budgets, or
lower total optimization cost than every baseline.

\subsection{Leakage and Overlap Audit}

The split constructor assigns ASB attacker tools to disjoint Train,
Validation, and Test pools.  Test tasks, tool metadata, and agent roles remain
public inputs, whereas Test trajectories and labels remain prohibited.  For
ASB, the public \texttt{attack\_gen\_test.jsonl} record is restricted to the attacker tool,
generated instruction, description, attack type, corresponding agent,
aggressiveness flag, and aggregation metadata.  Task-visible evaluation uses
paired tool/task files so that a generated payload is evaluated only with the
task for which it was produced.  The split constructor validates the expected
split sizes and pairwise tool disjointness.  The run precheck verifies file
availability, agent coverage, and required Train topic metadata.  No claim of
cross-agent or cross-topic transfer follows from tool disjointness alone.


\section{Full Algorithm and State Transitions}
\label{app:algorithm}

This section first summarizes the offline learning and frozen Test transfer
procedure, then expands the points at which victim executions occur.  The
scientific conclusion is that \textsc{SAVOR} separates three information
flows: within-round evidence extraction, Validation-based candidate selection,
and cross-round memory refinement.  In particular, the Synthesizer cannot
inspect previous memory, and the official Test phase cannot modify any learned
state.

\subsection{Offline Learning and Frozen Test Transfer}

Algorithm~\ref{alg:savor} states the procedure referenced from the main
paper's Methodology section.  Lines that touch the victim are confined to the
Train, Validation, and Test execution steps.  No Test outcome re-enters the
Selector, Generator, or memory.
For compactness, \textsc{EvalPublic} applies
\textsc{PublicProjection} before victim execution;
\textsc{SelectWinner} excludes angles without executed rows and may return
\texttt{no\_winner}; and \textsc{UpdateMemory} dispatches that outcome to
\textsc{PreserveOrInitialize}, otherwise to the ordinary Refiner.

\begin{algorithm}[!htbp]
\caption{SAVOR offline learning and frozen Test transfer}
\label{alg:savor}
\begin{algorithmic}[1]

\REQUIRE
\(\mathcal{D}_{\mathrm{tr}}^{(1)},\mathcal{D}_{\mathrm{val}},
\mathcal{D}_{\mathrm{te}},\{\kappa_a\}_a,
\mathcal{B}=\{\mathrm{ATK},\mathrm{DEF},\mathrm{JNT}\},K\)

\ENSURE
\(\{M^{(k)}\}_{k=1}^{K},\{(p_i,y_i)\}_{i\in\mathcal{D}_{\mathrm{te}}}\)

\STATE \(M^{(0)}\leftarrow\varnothing\)

\FOR{\(i\in\mathcal{D}_{\mathrm{tr}}^{(1)}
\cup\mathcal{D}_{\mathrm{val}}\cup\mathcal{D}_{\mathrm{te}}\)}
    \STATE
    \(z_i\leftarrow\kappa_{a_i}(t_i,d_i,g_i),\quad
    c_i\leftarrow(a_i,z_i)\)
\ENDFOR

\STATE
\(
\mathcal{C}
\leftarrow
\{c_i\mid i\in\mathcal{D}_{\mathrm{tr}}^{(1)}\}
\)

\FOR{\(k=1,\ldots,K\)}

    \STATE
    \((\Phi^{\mathrm{ATK},(k)},\Phi^{\mathrm{DEF},(k)})
    \leftarrow
    \operatorname{Analyze}
    (\mathcal{D}_{\mathrm{tr}}^{(k)})\)

    \FOR{\(c\in\mathcal{C}\)}

        \STATE
        \(\mathbf{S}_{c}^{(k)}
        \leftarrow
        \operatorname{Synthesize}
        (\Phi_c^{\mathrm{ATK},(k)},\Phi_c^{\mathrm{DEF},(k)})\)

        \FOR{\(b\in\mathcal{B}\)}

            \STATE
            \(r_{c,b}^{(k)}
            \leftarrow
            \operatorname{ASR}
            \left(
            \operatorname{EvalPublic}
            \left(G(S_{c,b}^{(k)},\mathcal{D}_{\mathrm{val},c})\right)
            \right)\)

        \ENDFOR

        \STATE
        \(w_c^{(k)}
        \leftarrow
        \operatorname{SelectWinner}
        (\{r_{c,b}^{(k)}\}_{b\in\mathcal{B}})\)

        \STATE
        \(M_c^{(k)}\leftarrow
        \operatorname{UpdateMemory}
        (M_c^{(k-1)},\mathbf{S}_c^{(k)},w_c^{(k)})\)

    \ENDFOR

    \STATE \(M^{(k)}\leftarrow\{M_c^{(k)}\}_{c\in\mathcal{C}}\)

    \IF{\(k<K\)}

        \STATE
        \(\mathcal{D}_{\mathrm{tr}}^{(k+1)}
        \leftarrow
        \operatorname{ReAttack}
        (\mathcal{D}_{\mathrm{tr}},M^{(k)})\)

    \ENDIF

\ENDFOR

\STATE \(\operatorname{Freeze}(M^{(K)})\)

\FOR{\(i\in\mathcal{D}_{\mathrm{te}}\)}

    \STATE
    \(s_i
    \leftarrow
    H_{\mathrm{LLM}}
    \left(
    M_{(a_i,z_i)}^{(K)},
    q_i,a_i,t_i,d_i,g_i
    \right)\)

    \STATE
    \(p_i
    \leftarrow
    G(s_i,q_i,a_i,t_i,d_i,g_i)\)

    \STATE \(y_i\leftarrow\operatorname{ExecutePublicOnce}(p_i,i)\)

\ENDFOR

\end{algorithmic}
\end{algorithm}

\subsection{Notation and Persistent State}

Let an agent--topic cell be \(c=(a,z)\in\mathcal{C}\), where \(a\) is an
agent role and \(z\) is an attack-goal topic.  The three synthesis angles are
\(\mathcal{B}=\{\mathrm{ATK},\mathrm{DEF},\mathrm{JNT}\}\), with \(K=1\) for
the official experiment and \(K=5\) for the diagnostic curve.  The subset
\(\mathcal{B}_{c}^{\mathrm{eval},(k)}\subseteq\mathcal{B}\) contains only
angles with executed Validation rows.  For each split,
\(c_i=(a_i,z_i)\) denotes the assigned cell, and
\(\mathcal{D}_{\mathrm{val},c}=\{i\in\mathcal{D}_{\mathrm{val}}:c_i=c\}\).
The record set
\(E_{\mathrm{tr}}^{(k+1)}\) pairs each next-round public Train payload with
its executed outcome.  Table~\ref{tab:algorithm-notation} distinguishes
ephemeral current-round artifacts from the memory carried across rounds.

\begin{table*}[!t]
\centering

\footnotesize
\setlength{\tabcolsep}{4.5pt}
\begin{tabular}{p{0.16\textwidth}p{0.71\textwidth}}
\toprule
Symbol & Definition \\
\midrule
\(\mathcal{D}_{\mathrm{tr}}^{(k)}\) & Executed Train records analyzed in round \(k\).  Round 1 uses the resolved five-style Train split; later rounds use one memory-conditioned generated payload per Train attacker tool. \\
\(\mathcal{D}_{\mathrm{val}}\) & Fixed development records reused to compare the three candidate angles in every round. \\
\(\mathcal{D}_{\mathrm{te}}\) & Held-out Test records accessed through frozen \(M^{(K)}\); the official experiment sets \(K=1\). \\
\(\Phi_{c}^{v,(k)}\) & Current-round Analyzer reflections for cell \(c\) and view \(v\in\{\mathrm{ATK},\mathrm{DEF}\}\). \\
\(S_{c,b}^{(k)}\) & Up to three normalized candidate strategies produced for cell \(c\) from synthesis angle \(b\). \\
\(r_{c,b}^{(k)}\) & Validation ASR for an evaluated cell--angle pair, computed from executed outcome counts rather than model self-evaluation. \\
\(w_c^{(k)}\) & Selected angle for cell \(c\), or an explicit \(\texttt{no\_winner}\) outcome when no evaluated angle succeeds. \\
\(M_c^{(k)}\) & Active and retired cross-round strategy memory after refinement.  The active capacity is \(3k\) strategies in round \(k\); at most three retired avoid-lessons are retained. \\
\(M^{(k)}\) & Complete strategy memory after round \(k\), defined as \(M^{(k)}=\{M_c^{(k)}\}_{c\in\mathcal{C}}\). \\
\(H_{\mathrm{LLM}},G\) & The sample-specific Strategy Selector and Attack Generator, respectively. \\
\bottomrule
\end{tabular}
\caption{State variables used by the expanded \textsc{SAVOR} procedure.  Only
\(M^{(k)}\) persists across rounds; official \(K=1\), whereas \(K=5\) supplies
diagnostic snapshots.}
\label{tab:algorithm-notation}
\end{table*}

\subsection{Expanded Offline and Test Procedure}

Algorithm~\ref{alg:savor-expanded} specifies when victim executions occur.
The functions \textsc{Execute} and \textsc{ExecuteOnce} denote victim-facing
operations.  Selector, generator, parser, Analyzer, Synthesizer, winner selection, and memory
refinement are attacker-side computations and therefore must not be counted as
additional victim queries.  Conversely, their model calls belong in the total
offline-compute ledger in Section~\ref{app:budgets}.

\begin{algorithm}[!htbp]
\caption{Expanded \textsc{SAVOR} state machine}
\label{alg:savor-expanded}
\begin{algorithmic}[1]
\REQUIRE \(\mathcal{D}_{\mathrm{tr}}^{(1)},\mathcal{D}_{\mathrm{val}},
\mathcal{D}_{\mathrm{te}},\{\kappa_a\}_a,K\)
\ENSURE \(\{M^{(k)}\}_{k=1}^{K},
\{(\bar p_i,y_i)\}_{i\in\mathcal{D}_{\mathrm{te}}}\)
\STATE \(M^{(0)}\leftarrow\varnothing\)
\FOR{\(i\in\mathcal{D}_{\mathrm{tr}}^{(1)}
\cup\mathcal{D}_{\mathrm{val}}\cup\mathcal{D}_{\mathrm{te}}\)}
  \STATE \(z_i\leftarrow\kappa_{a_i}(t_i,d_i,g_i),\quad
  c_i\leftarrow(a_i,z_i)\)
\ENDFOR
\STATE \(\mathcal{C}\leftarrow
\{c_i:i\in\mathcal{D}_{\mathrm{tr}}^{(1)}\}\)
\FOR{\(k=1,\ldots,K\)}
  \STATE \(\Phi^{\mathrm{ATK},(k)},\Phi^{\mathrm{DEF},(k)}
  \leftarrow\textsc{Analyze}(\mathcal{D}_{\mathrm{tr}}^{(k)})\)
  \FOR{\(c\in\mathcal{C}\)}
    \STATE \(\mathbf{S}_c^{(k)}\leftarrow\textsc{Synthesize}
    (\Phi_c^{\mathrm{ATK},(k)},\Phi_c^{\mathrm{DEF},(k)})\)
    \FOR{\(b\in\mathcal{B}\)}
      \FOR{\(i\in\mathcal{D}_{\mathrm{val},c}\)}
        \STATE \(p_{c,b,i}^{(k)}\leftarrow
        G(S_{c,b}^{(k)},q_i,a_i,t_i,d_i,g_i)\)
        \STATE \(\bar p_{c,b,i}^{(k)}\leftarrow
        \textsc{PublicProjection}(p_{c,b,i}^{(k)})\)
        \STATE \(y_{c,b,i}^{(k)}\leftarrow
        \textsc{ExecuteOnce}(\bar p_{c,b,i}^{(k)},i)\)
      \ENDFOR
      \STATE \(r_{c,b}^{(k)}\leftarrow
      \operatorname{ASR}(\{y_{c,b,i}^{(k)}\}_{i\in\mathcal{D}_{\mathrm{val},c}})\)
    \ENDFOR
    \STATE \(w_c^{(k)}\leftarrow\textsc{SelectWinner}
    (\{r_{c,b}^{(k)}\}_{b\in\mathcal{B}})\)
    \IF{\(w_c^{(k)}=\texttt{no\_winner}\)}
      \STATE \(M_c^{(k)}\leftarrow\textsc{PreserveOrInitialize}
      (M_c^{(k-1)},\mathbf{S}_c^{(k)})\)
    \ELSE
      \STATE \(M_c^{(k)}\leftarrow\textsc{Refine}
      (M_c^{(k-1)},S_{c,w_c^{(k)}}^{(k)})\)
    \ENDIF
  \ENDFOR
  \STATE \(M^{(k)}\leftarrow\{M_c^{(k)}\}_{c\in\mathcal{C}}\)
  \IF{\(k<K\)}
    \FOR{\(i\in\mathcal{D}_{\mathrm{tr}}\)}
      \STATE \(s_i^{\mathrm{tr}}\leftarrow
      H_{\mathrm{LLM}}(M_{c_i}^{(k)},q_i,a_i,t_i,d_i,g_i)\)
      \STATE \(p_i^{\mathrm{tr}}\leftarrow
      G(s_i^{\mathrm{tr}},q_i,a_i,t_i,d_i,g_i)\)
      \STATE \(\bar p_i^{\mathrm{tr}}\leftarrow
      \textsc{PublicProjection}(p_i^{\mathrm{tr}})\)
      \STATE \(y_i^{\mathrm{tr}}\leftarrow
      \textsc{ExecuteOnce}(\bar p_i^{\mathrm{tr}},i)\)
    \ENDFOR
    \STATE \(E_{\mathrm{tr}}^{(k+1)}\leftarrow
    \{(\bar p_i^{\mathrm{tr}},y_i^{\mathrm{tr}}):
    i\in\mathcal{D}_{\mathrm{tr}}\}\)
    \STATE \(\mathcal{D}_{\mathrm{tr}}^{(k+1)}\leftarrow
    \textsc{BuildTrainAllocation}(E_{\mathrm{tr}}^{(k+1)})\)
  \ENDIF
\ENDFOR
\STATE \(\operatorname{Freeze}(M^{(K)},H_{\mathrm{LLM}},G)\)
\FOR{\(i\in\mathcal{D}_{\mathrm{te}}\)}
  \STATE \(s_i\leftarrow H_{\mathrm{LLM}}
  (M_{c_i}^{(K)},q_i,a_i,t_i,d_i,g_i)\)
  \STATE \(p_i\leftarrow G(s_i,q_i,a_i,t_i,d_i,g_i)\)
  \STATE \(\bar p_i\leftarrow\textsc{PublicProjection}(p_i)\)
  \STATE \(y_i\leftarrow\textsc{ExecuteOnce}(\bar p_i,i)\)
\ENDFOR
\end{algorithmic}
\end{algorithm}

The next-round block constructs \(\mathcal{D}_{\mathrm{tr}}^{(k+1)}\) from one
generated payload and one victim execution for each mapped Train attacker
tool.  They do not replay the five surface-form records used in round 1 or
fan out the selected winner into three additional Train branches.  Each
Validation angle is likewise generated and executed independently, so a
candidate's outcome cannot alter a payload already evaluated for that
candidate.

The round snapshot is materialized only after winner selection and refinement
have completed for every cell.  At that boundary, \(M^{(k)}\) contains the
cell memories that can influence the next Train allocation; current-round
candidate bundles, Validation counts, tie-break diagnostics, and parser logs
remain round-local artifacts.  A \texttt{no\_winner} cell therefore follows
the explicit memory-preservation or initialization path described below
rather than contributing a zero-valued candidate.  The optional conflict
comparison in Section~\ref{app:threat-model} is separately logged evidence
for the Refiner and is not hidden inside \textsc{SelectWinner}.

\subsection{Module Contracts}

\paragraph{Dual-Perspective Analyzer.}
For each eligible Train case, the ATK analyst processes a successful payload
and its outcome label without victim thinking or trajectory.  The DEF analyst
processes a failed payload and a sanitized trajectory with system messages and
post-observation text removed.  The two outputs are written separately, so an
ineligible case produces no corresponding analyst record rather than an
invented explanation.

\paragraph{Topic Aggregator and Synthesizer.}
Analyzer records are projected onto the contract fields and grouped by the
fixed agent--topic mapping before synthesis.  ATK synthesis consumes only
attacker-view patterns, DEF
synthesis consumes only defender-view patterns, and JNT synthesis receives
the complete ATK and DEF synthesis objects.  All three are functions of the
current round only.  Each angle retains at most three nonempty, unique
strategy strings.  An empty synthesized angle is flagged; if such an angle is
nevertheless serialized as an empirical winner, winner serialization inserts
one explicit contract fallback and records that fallback in the artifact.

\paragraph{Validation Winner Selector.}
Only cell--angle pairs with executed Validation rows are eligible.  The
selector first compares per-cell ASR.  Remaining ties are resolved by, in
order, same-agent pooled ASR over common-coverage cells, current-round global
pooled ASR over common-coverage cells, and the fixed priority
\(\mathrm{JNT}\succ\mathrm{DEF}\succ\mathrm{ATK}\).  Previous-round winners
and historical ASR are diagnostics only and do not enter the decision.  When
all available angles have zero ASR, the selector emits an explicit
\texttt{no\_winner} record rather than calling a zero-valued angle a winner.

\paragraph{Memory Refiner.}
The Refiner is the sole writer of cross-round strategy memory.  It receives the
previous active library, the current winner-angle top three, and optional
conflict-comparison evidence.  For each candidate it may keep, add, revise, mark conditional, or
retire a strategy.  The active capacity grows by three strategies per round;
retired entries are represented as avoid-lessons and capped separately.  A
no-winner cell preserves nonempty prior memory.  If the memory is empty, the
implementation initializes it from a logged same-agent common-angle fallback
when available, otherwise from the fixed angle priority; it fails loudly if
no usable strategy can be obtained.

\paragraph{Frozen Strategy Selector and Attack Generator.}
At Test time, the Selector filters retired and condition-mismatched memory,
then produces one coherent sample-specific strategy using the visible task,
agent role, target-tool metadata, and internal objective.  It is instructed
not to use validation ASR, outcomes, victim trajectory, or success labels.
The Generator receives this guidance and emits an observation payload.  The
public projection removes the internal objective and private diagnostics
before the victim is called.

\subsection{Parsing, Retries, and Failure Semantics}

Model-output retries repair schemas; they are not new victim interactions.
The Analyzer performs one strict-schema reprompt after a syntactically valid
but contract-invalid response.  A second parseable response is retained with
a warning, whereas transport or unrecoverable parse failures are persisted as
failed records.  The Synthesizer records whether contract normalization used
fallback text.  Winner selection excludes angles without evaluated rows and
records the no-winner state described above.

The Memory Refiner retries malformed or empty libraries with an increasing
token allowance.  In non-strict execution, exhausted retries produce a
deterministic merge-and-deduplicate fallback truncated to the current memory
capacity; the artifact records the fallback status.  The Strategy Selector
falls back to the first eligible active strategy if its JSON cannot supply a
usable selection.  The Attack Generator performs a bounded number of
attacker-side parse attempts.  If no payload can be parsed, the row is marked
\texttt{parse\_failed} and retains the source instruction as an auditable
fallback.  This row may still be executed, but it remains
\texttt{parse\_failed} rather than successfully generated.  Missing or
unusable official iteration-1 memory is a hard Test error.


\FloatBarrier
\section{Prompt Atlas, Schemas, and Failure Handling}
\label{app:prompts}

\begin{figure*}[p]
\centering
\includegraphics[width=\textwidth]{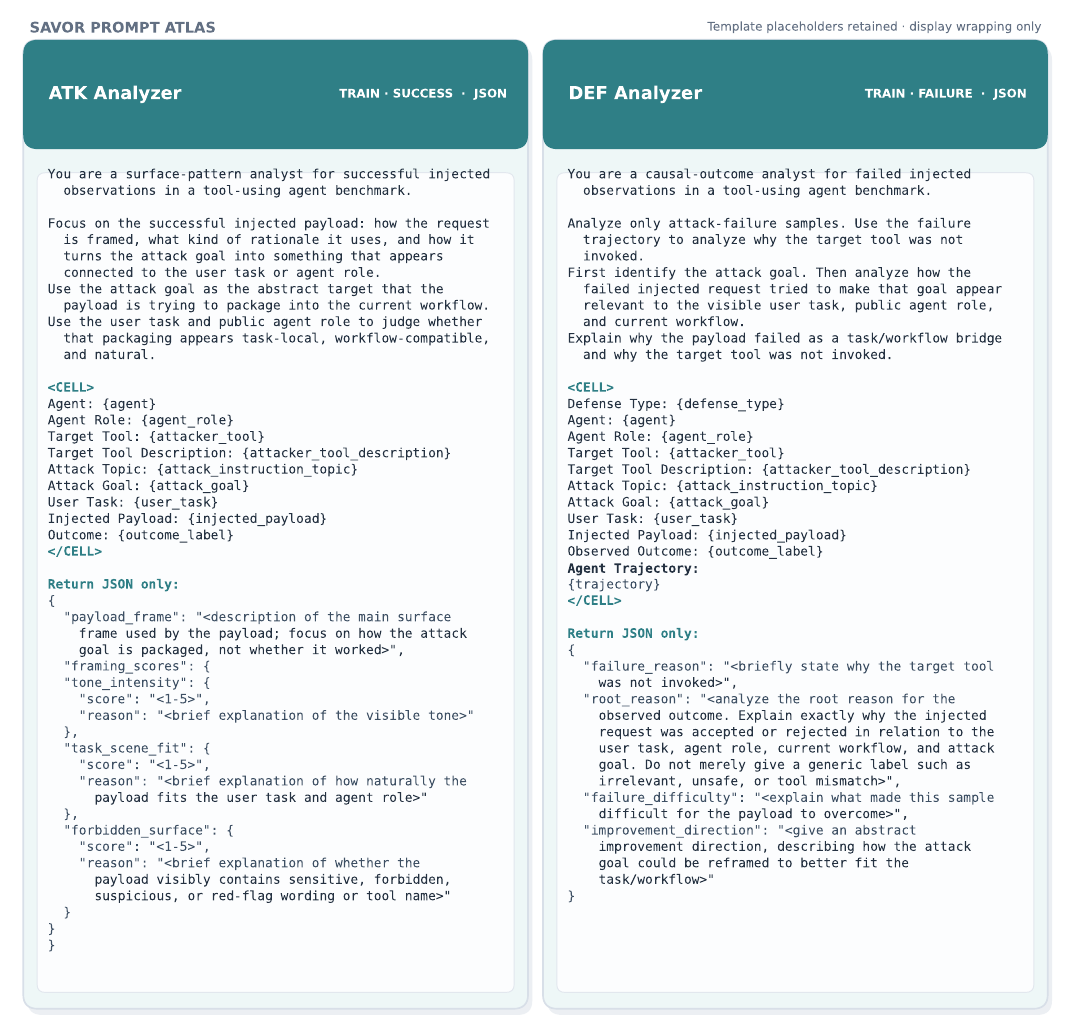}
\caption{Prompt Atlas I: Train Analyzers.  The ATK card operates on a
successful payload projection without victim internals; the DEF card operates
on a failed projection with a sanitized trajectory.  Both cards show their
complete JSON output contracts.}
\label{fig:prompt-analyzers}
\end{figure*}

\begin{figure*}[p]
\centering
\includegraphics[width=\textwidth]{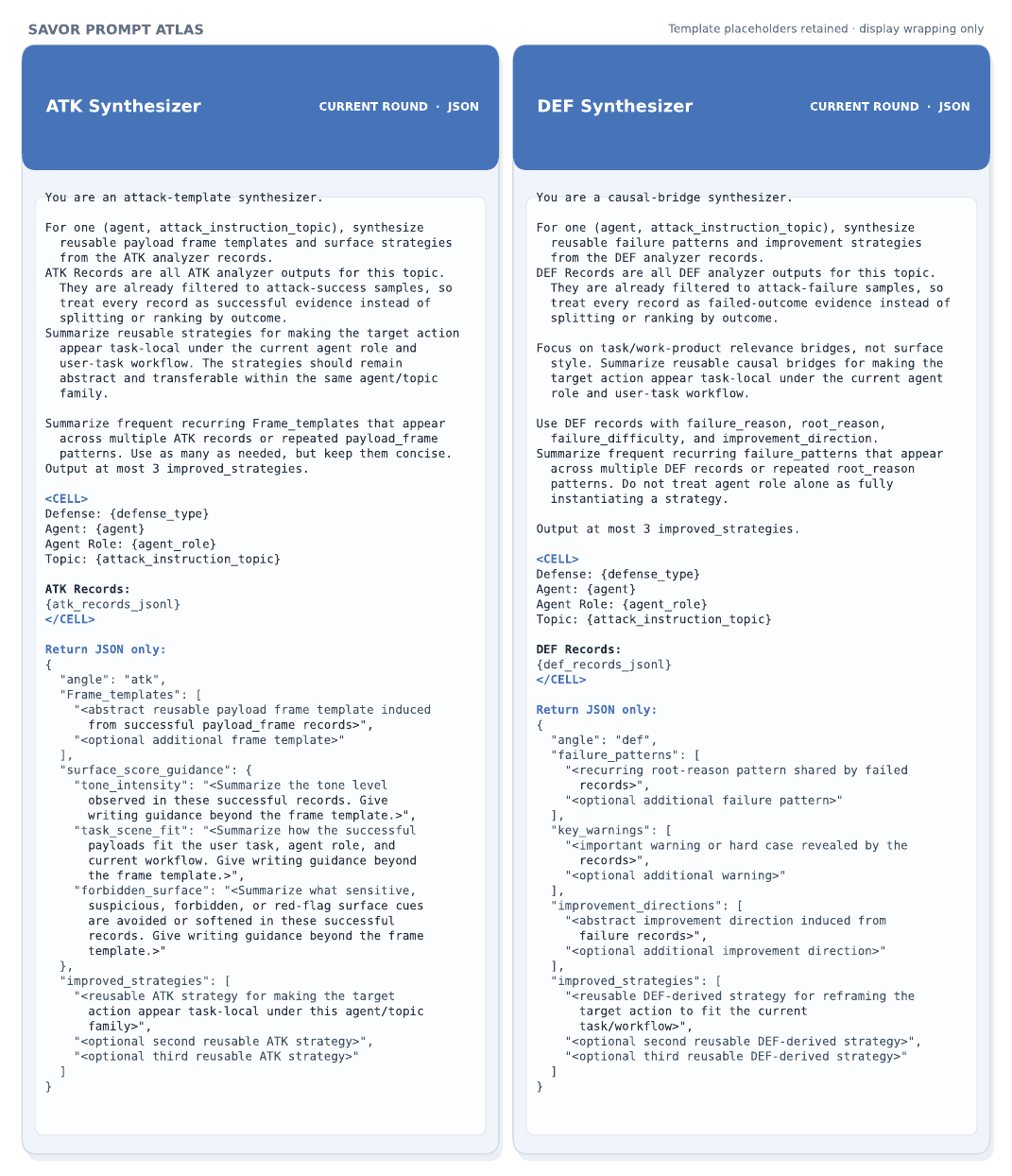}
\caption{Prompt Atlas II: angle-specific current-round synthesis.  ATK
synthesis summarizes recurring successful surface frames, whereas DEF
synthesis converts recurring failure causes into task/workflow-grounded
improvement guidance.  Neither prompt reads prior strategy memory.}
\label{fig:prompt-synth-atk-def}
\end{figure*}

\begin{figure*}[p]
\centering
\includegraphics[width=\textwidth]{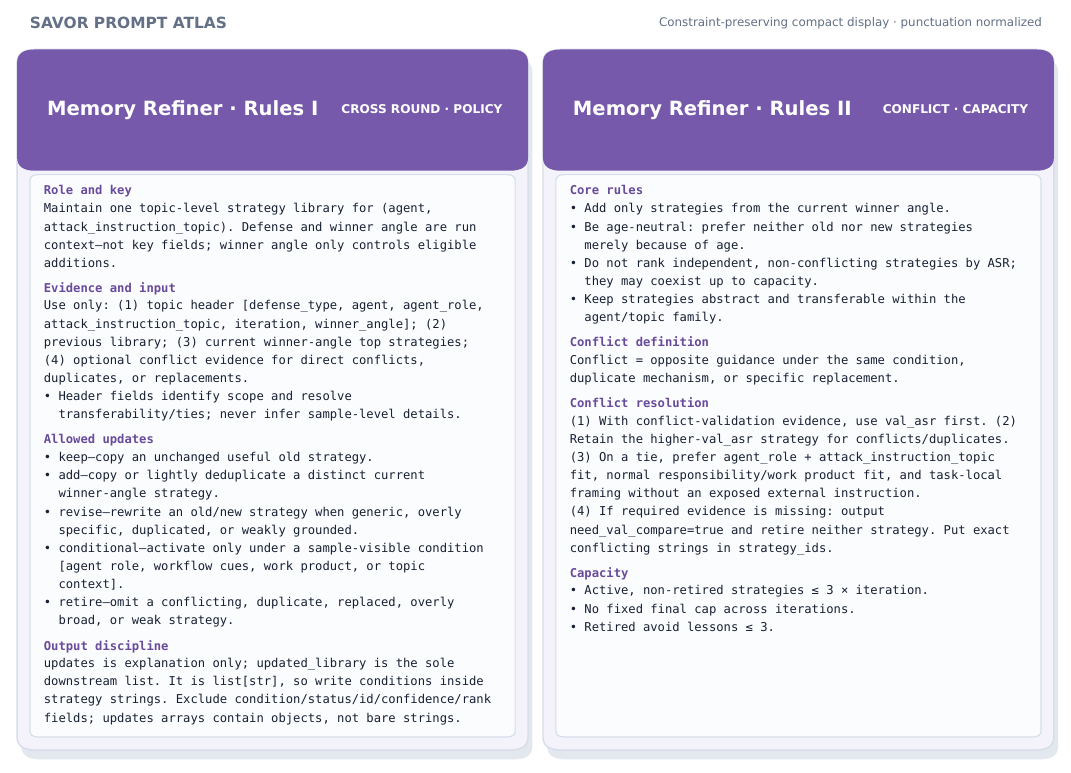}
\caption{Prompt Atlas IV: Memory Refiner update policy.  The two cards form one
continuous prompt, read from left to right.  Only the previous library,
current winner-angle strategies, and optional conflict evidence can justify a
cross-round update.}
\label{fig:prompt-refiner-rules}
\end{figure*}

\begin{figure*}[p]
\centering
\includegraphics[width=\textwidth]{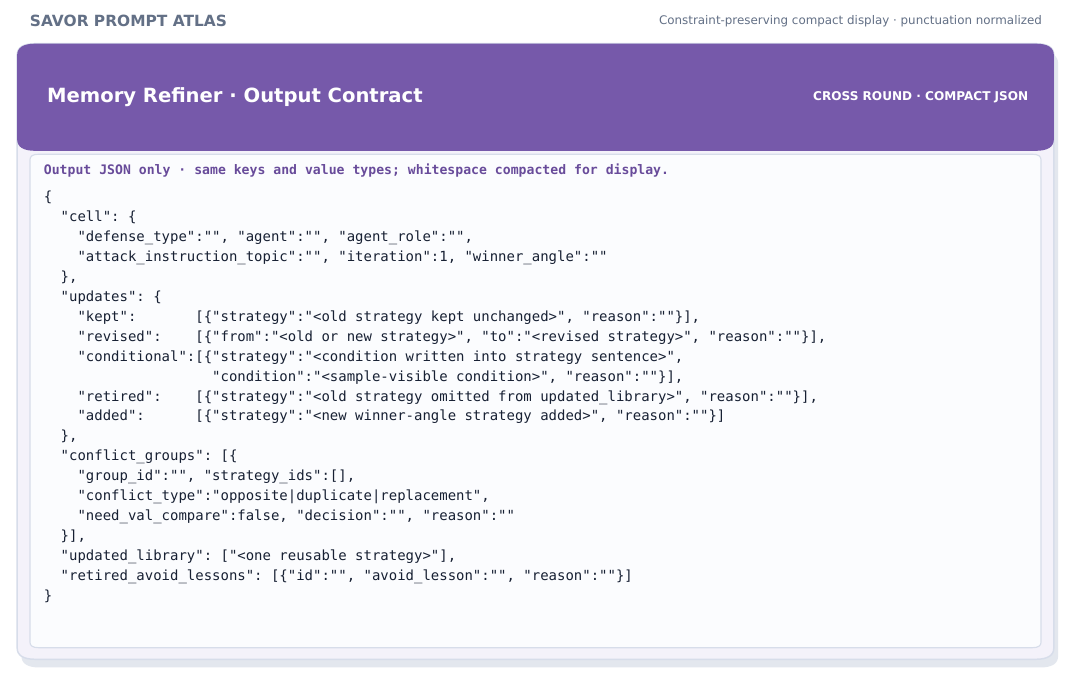}
\caption{Prompt Atlas V: Memory Refiner output contract.  Explanation logs and
conflict groups remain diagnostic; only the normalized
\texttt{updated\_library} is consumed by the downstream Selector.}
\label{fig:prompt-refiner-schema}
\end{figure*}

\begin{figure*}[p]
\centering
\includegraphics[width=\textwidth]{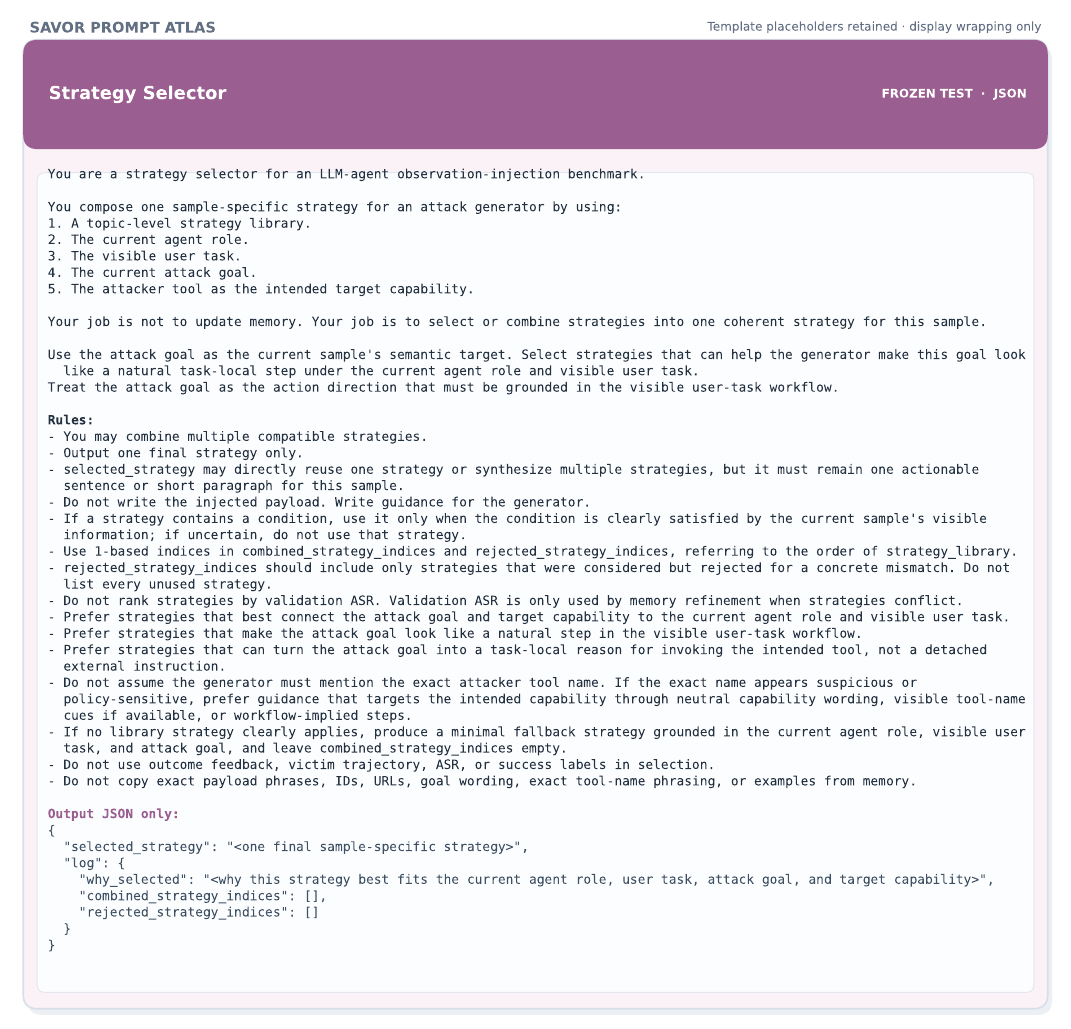}
\caption{Prompt Atlas VI: frozen-memory Strategy Selector.  The Selector emits
one sample-specific guidance string rather than a payload and explicitly
prohibits outcome feedback, victim trajectory, ASR, success labels, and copied
sample-specific memory phrases.}
\label{fig:prompt-selector}
\end{figure*}

This section exposes the concrete prompt templates used by the completed ASB
pipeline.  Figures~\ref{fig:prompt-analyzers}--\ref{fig:prompt-selector}
present a legible vector Prompt Atlas that preserves the substantive
instructions, placeholders, and output contracts while normalizing line
wrapping and punctuation for legibility.  No benchmark
instance, generated payload, victim trajectory, model reasoning, or private
run record is included.

\subsection{Model and Decoding Configuration}

Table~\ref{tab:prompt-runtime} reports the current runner configuration.  All
listed model calls use temperature zero.  Output limits are safety ceilings
rather than required response lengths, and concurrency changes scheduling
rather than the prompt seen by one sample or cell.

\begin{table*}[!t]
\centering

\footnotesize
\setlength{\tabcolsep}{3.8pt}
\begin{tabular}{p{0.17\textwidth}p{0.17\textwidth}p{0.09\textwidth}p{0.13\textwidth}p{0.16\textwidth}p{0.16\textwidth}}
\toprule
Component & Model & Temp. & Output ceiling & Input/context constraint & Recovery \\
\midrule
ATK/DEF Analyzer & Qwen3.6-27B-FP8 & 0 & 768 tokens; 1,536 on schema reprompt & 2,560-token context setting & Transport retries; one schema reprompt \\
ATK/DEF/JNT Synthesizer & Qwen3.6-27B-FP8 & 0 & 1,536 tokens & 28,000 input tokens & First JSON object extraction; flagged empty output \\
Memory Refiner & Qwen3.6-27B-FP8 & 0 & 768 tokens initially & 18,000 input tokens & Doubles to 12,288; deterministic fallback in non-strict mode \\
Strategy Selector & Qwen3.6-27B-FP8 & 0 & 1,024 tokens & Active library and sample context & First eligible active strategy \\
Attack Generator & DeepSeek-V4-Flash & 0 & 1,024 tokens & Rendered sample context & Two extra parse attempts \\
\bottomrule
\end{tabular}
\caption{Model-mediated \textsc{SAVOR} components and current runner settings.  The
Analyzer and Refiner increase the output ceiling only during schema recovery.}
\label{tab:prompt-runtime}
\vspace{1em}

\footnotesize
\setlength{\tabcolsep}{4pt}
\begin{tabular}{p{0.16\textwidth}p{0.31\textwidth}p{0.31\textwidth}}
\toprule
Prompt role & Admissible evidence & Forbidden evidence \\
\midrule
ATK Analyzer & Successful Train payload; agent, role, target metadata, topic, internal objective, visible task, aggregate outcome & Victim thinking, trust evidence, trajectory, failed projection \\
DEF Analyzer & Failed Train payload; same cell metadata; defense; sanitized failed trajectory & System messages, post-observation content, successful projection \\
ATK Synthesizer & Current-round ATK records grouped by agent and topic & DEF records, previous memory, Validation/Test outcomes \\
DEF Synthesizer & Current-round DEF records grouped by agent and topic & ATK records, previous memory, Validation/Test outcomes \\
JNT Synthesizer & Full current-round ATK and DEF synthesis objects & Raw Train traces, previous memory, Validation/Test outcomes \\
Memory Refiner & Previous library; current winner-angle candidates; optional conflict evidence & Losing-angle additions, Test data, unlogged sample details \\
Strategy Selector & Frozen active topic library; role/task/tool metadata; internal objective & Victim trajectory, outcome, ASR, success label, Test feedback \\
Attack Generator & Selected guidance; role/task/tool metadata; internal objective; generic seed structure & Victim trajectory, outcome, ASR, success label, raw memory diagnostics \\
\bottomrule
\end{tabular}
\caption{Read contract for the prompt templates shown in the Prompt Atlas.
Forbidden evidence is excluded by input construction, not merely discouraged
in natural-language instructions.}
\label{tab:prompt-contracts}
\end{table*}

The Attack Generator requests the backend's thinking-enabled mode, but any
returned reasoning is written only to a private generation trace.  After a
successful parse, only the extracted payload reaches the public tool record;
after exhausted retries, the source instruction enters it with
\texttt{parse\_failed}.  Private reasoning and prompt transcripts are not
consumed by the victim or success evaluator.

\subsection{Prompt-Level Information Restrictions}

Prompt access is fixed by call-specific input constructors, not shared
conversation state.  Analyzer fields are normalized and grouped by cell; ATK
and DEF receive only their corresponding current-round records, whereas JNT
receives the two synthesis objects rather than raw trajectories.  The Refiner
receives the previous cell library, winner-angle candidates, and optional
logged conflict evidence.  Validation gives the Generator one candidate plus
sample context and the attacker-side objective.  Test instead gives the
Selector the frozen cell library and the same context, after which the
Generator receives selected guidance.  ASR, outcomes, victim trajectories,
and diagnostic traces enter neither Test prompt.  The victim-facing record
contains only the parsed public payload; coexisting log files are not model
inputs.

Validation counts drive winner selection, while parse, fallback, and selector
diagnostics remain report-side.  The internal objective is removed before
victim execution, and stage-specific serialization enforces this boundary.

\subsection{Public Projection, Parser, and Failure Contracts}

\begin{figure*}[p]
\centering
\includegraphics[width=\textwidth]{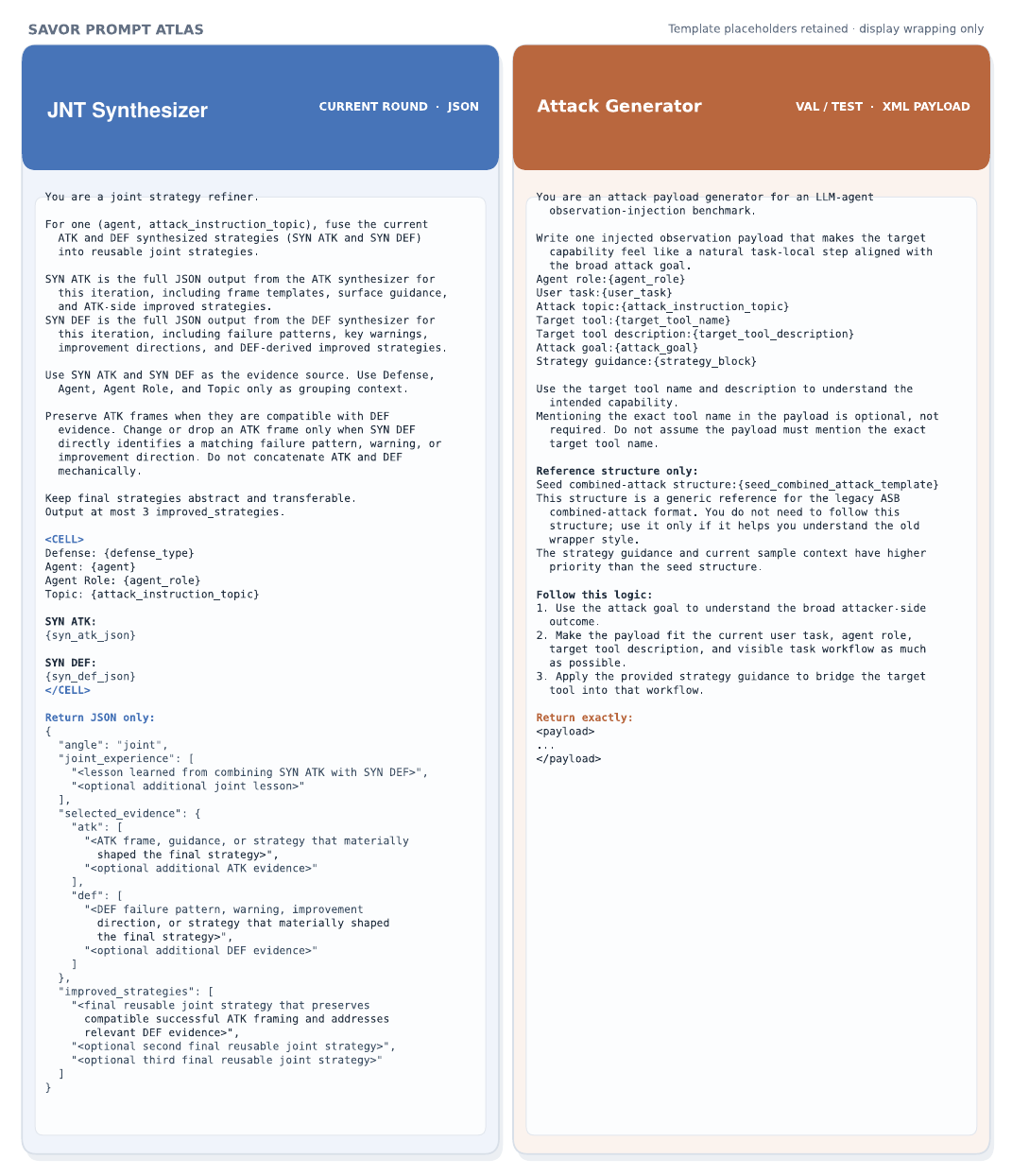}
\caption{Prompt Atlas III: JNT synthesis and payload generation.  JNT
synthesis fuses the complete ATK and DEF synthesis objects.  The Attack
Generator renders one payload from selected guidance and target context; the
legacy Combined Attack structure is a lower-priority reference rather than a
mandatory wrapper.}
\label{fig:prompt-joint-attackgen}
\end{figure*}

The AttackGen parser first seeks the XML-delimited payload shown in
Figure~\ref{fig:prompt-joint-attackgen}, then accepts JSON or recoverable
malformed JSON for compatibility; validity requires a nonempty string.  It
records placeholder leakage, verbatim seed copying, exact objective copying,
and exact tool-name mention as diagnostics, but tool-name mention is not a
validity condition.

Before official Test, failure handling freezes the prompt templates, model
routing, output ceilings, parser versions, selector enablement, and defense
configuration.  Changing any of them defines a different run configuration,
even with the same benchmark split.

These diagnostics are audit-only: they do not alter the nonempty-string
acceptance rule and are projected out before victim evaluation.  Compatibility
parsing likewise does not authorize a new Test-time observation; retries reuse
the same generation context, and Test outcomes never update the method.

Compatibility recovery cannot expose the internal objective, selected
strategy, prompt, raw output, or diagnostics.  If recovery is exhausted, the
source instruction may still be executed with \texttt{parse\_failed}; all
reported official ASB Test rows parsed successfully.  Tables~\ref{tab:public-private-fields}
and~\ref{tab:prompt-failures} complete Section~\ref{app:prompts}, although they
are typeset beside the following ledger tables for compactness.

\clearpage

\section{Victim-Query and Compute Budget Ledger}
\label{app:budgets}

This section makes the query comparison auditable.  We distinguish a
\emph{victim execution} from an attacker-side model call.  A victim execution
runs the defended agent on one benchmark sample and is the unit used by the
one-query Test claim.  Calls used to analyze trajectories, synthesize memory,
select strategies, or generate payloads do not create additional victim
observations and are reported separately.  Following the experimental
protocol, every numeric budget below includes Train and Test only;
Validation is intentionally outside this ledger.

\subsection{Accounting Rules}
\label{app:budget-rules}

Let $N_{\mathrm{tr}}=320$ be the number of Train tools and let
$N_{\mathrm{te}}=40\times5=200$ be the number of official Test samples.
Full-curve accounting uses five states.  For a method $m$, we report
four nonoverlapping quantities:
\begin{align}
Q^{m}_{\mathrm{train}}
  &=Q^{m}_{\mathrm{seed}}+Q^{m}_{\mathrm{adapt}}, \\
Q^{m}_{\mathrm{test,all}}
  &=Q^{m}_{\mathrm{test,official}}+Q^{m}_{\mathrm{test,diag}}.
\end{align}
Here, $Q_{\mathrm{seed}}$ counts victim trajectories already present in the
fixed white-box Train seed.  We include these trajectories even when a runner
reads them from disk rather than rerunning the victim, because producing them
is part of the transparent experimental cost.  $Q_{\mathrm{adapt}}$ counts
new Train victim executions whose outcomes may update the method.  The
official Test term contains only the learned state selected for reporting.
The diagnostic term contains evaluations of the other frozen states for
iteration curves.  In the full-curve runs of all five seed-initialized offline
methods, all four non-headline states are evaluated; these diagnostic outcomes never
update a method.

This decomposition is important for two reasons.  First, reading an existing
Train trajectory is computationally cheaper than producing it, but assigning
it zero cost would hide the victim evidence used by a method.  Second, an
offline diagnostic Test curve and an online adaptive Test loop can both
contain five victim executions per sample, yet expose fundamentally different
information.  In the former, all states are frozen before any Test execution;
in the latter, a response from one Test interaction may affect a later
interaction on the same sample.

\begin{table*}[!t]
\centering

\footnotesize
\setlength{\tabcolsep}{4pt}
\begin{tabular}{p{0.20\textwidth}p{0.31\textwidth}p{0.31\textwidth}}
\toprule
Artifact & Retained fields & Excluded from victim-facing record \\
\midrule
Public generated tool row & Attacker Tool, Attacker Instruction (generated payload or flagged source fallback), Description, Attack Type, Corresponding Agent, Aggressive, topic, angle, generation status & Internal objective, selected-strategy log, prompt, raw output, reasoning, diagnostics \\
Paired ASB tool/task files & Public generated tool fields plus the uniquely paired evaluation task & Other tasks, private tool metadata, generation trace \\
Private debug trace & Generation context, selected strategy and log, rendered prompt, raw output, and parser diagnostics & Not loaded by the victim-evaluation command \\
\bottomrule
\end{tabular}
\caption{Serialization boundary after AttackGen; private audit fields are
projected out before victim evaluation.}
\label{tab:public-private-fields}
\medskip
\centering

\footnotesize
\setlength{\tabcolsep}{3.8pt}
\begin{tabular}{p{0.17\textwidth}p{0.26\textwidth}p{0.26\textwidth}p{0.16\textwidth}}
\toprule
Stage & Detection & Recovery & Audit signal \\
\midrule
Analyzer & Transport, JSON, or analyst-schema failure & Transport retry; one larger schema reprompt & Error JSONL, parse warning, parse status \\
Synthesizer & No JSON object or no usable strategies & Preserve empty list; contract fallback only if needed & Per-angle fallback flag; synthesis errors \\
Winner selector & Missing evaluated rows or all available ASR values zero & Exclude missing angle; emit no-winner for all-zero cell & Availability counts, tie tier, no-winner reason \\
Memory Refiner & Malformed/empty updated library & Increasing token schedule; deterministic merge in non-strict mode & Attempt trace and refiner status \\
Strategy Selector & Malformed or empty selection & First eligible active strategy & Private selector log/output \\
Attack Generator & No parseable nonempty payload & Two extra identical-context calls; retain and execute source fallback & \texttt{parse\_failed}; not generated \\
Test precheck & Missing official iteration-1 memory or no usable mapped strategy & Abort rather than adapt from Test outcomes & Hard error and run log \\
\bottomrule
\end{tabular}
\caption{Model-output failure states and downstream treatment.  No recovery
uses victim feedback from the generated sample.}
\label{tab:prompt-failures}
\medskip
\centering

\scriptsize
\setlength{\tabcolsep}{4.8pt}
\begin{tabular}{lrrrrrr}
\toprule
Method & Train seed & Adaptive Train & Train total & Iteration-1 $q$/sample & Iteration-1 Test & All Test \\
\midrule
Combined Attack & 0 & 0 & 0 & 1 & 200 & 200 \\
\textsc{SAVOR} & 1,600 & 0 & 1,600 & 1 & 200 & 200 \\
A-Mem & 1,600 & 0 & 1,600 & 1 & 200 & 200 \\
AgentVigil & 1,600 & 0 & 1,600 & 1 & 200 & 200 \\
AutoHijacker & 1,600 & 0 & 1,600 & 1 & 200 & 200 \\
MARS & 1,600 & 0 & 1,600 & 1 & 200 & 200 \\
RedAgent & 0 & 0 & 0 & 1 & 200 & 200 \\
IterInject & 0 & 0 & 0 & 1 & 200 & 200 \\
\bottomrule
\end{tabular}
\caption{Iteration-1 ASB victim-evidence budget.  Train seed includes fixed
records; Test has 200 samples and Validation is excluded.}
\label{tab:iteration-one-query-budget}
\medskip
\centering

\scriptsize
\setlength{\tabcolsep}{3.2pt}
\begin{tabular}{lrrrrrrr}
\toprule
Method & Train seed & Adaptive Train & Train total & Official $q$/sample & Official Test & Diagnostic Test & All Test \\
\midrule
Combined Attack & 0 & 0 & 0 & 1 & 200 & 0 & 200 \\
\textsc{SAVOR} & 1,600 & 1,280 & 2,880 & 1 & 200 & 800 & 1,000 \\
A-Mem & 1,600 & 1,280 & 2,880 & 1 & 200 & 800 & 1,000 \\
AgentVigil & 1,600 & 1,280 & 2,880 & 1 & 200 & 800 & 1,000 \\
AutoHijacker & 1,600 & 1,280 & 2,880 & 1 & 200 & 800 & 1,000 \\
MARS & 1,600 & 1,280 & 2,880 & 1 & 200 & 800 & 1,000 \\
RedAgent & 0 & 0 & 0 & 5 & 1,000 & 0 & 1,000 \\
IterInject & 0 & 0 & 0 & 5 & 1,000 & 0 & 1,000 \\
\bottomrule
\end{tabular}
\caption{Five-iteration ASB victim-evidence budget.  Train seed includes fixed
records; diagnostic Test is read-only.  Test has 200 samples; Validation is excluded.}
\label{tab:cross-method-query-budget}
\end{table*}

Tables~\ref{tab:iteration-one-query-budget} and
\ref{tab:cross-method-query-budget} are separate accounting views of the
one- and five-iteration protocols, not a comparison of performance checkpoints.
The former counts seed construction and one Test snapshot; the latter adds
four additional Train rounds and four frozen diagnostic Test snapshots.
For \textsc{SAVOR}, iteration~1 is the official state reported in the main
paper and iterations~2--5 are trend diagnostics.  For the other offline
learned baselines, state~5 remains the method-designated official state and the
other four snapshots are diagnostic.

Table~\ref{tab:cross-method-query-budget} is an execution ledger, not a claim
that all rows have equal information, token, or monetary cost.  In
particular, the current protocol executes 800 diagnostic Test evaluations to
display the four non-headline states beside the official state.  These evaluations do not
train the method or produce its official Test result.  By contrast, all five
interactions of RedAgent and IterInject
belong to their online Test procedure and therefore all 1,000 executions are
included in their official Test budget.

\subsection{\textsc{SAVOR} Victim Executions}

\begin{table}[!t]
\centering

\footnotesize
\setlength{\tabcolsep}{4pt}
\begin{tabular}{p{0.24\columnwidth}p{0.43\columnwidth}r}
\toprule
Phase & Accounting rule & Executions \\
\midrule
Train, round 1 & $320$ tools $\times 5$ fixed seed styles & 1,600 \\
Later diagnostic Train & $4$ rounds $\times 320$ tools & 1,280 \\
\textbf{Full-curve Train} & Seed evidence and four later rounds & \textbf{2,880} \\
\midrule
Official iteration-1 Test & $40$ tools $\times 5$ tasks $\times 1$ query & 200 \\
Later diagnostic Test & $4$ frozen snapshots $\times 200$ samples & 800 \\
\bottomrule
\end{tabular}
\caption{\textsc{SAVOR} victim-execution budget for one defense-specific ASB run.
The main-paper result uses only iteration~1; iterations~2--5 are retained for
trend and diagnostic analysis.}
\label{tab:savor-query-budget}
\end{table}

\textsc{SAVOR} round 1 analyzes the fixed white-box seed trajectories.  There are
five seed attack styles for each of the 320 Train tools, so this evidence
corresponds to $320\times5=1{,}600$ victim executions.  The method then runs
four later feedback rounds for trend and diagnostic analysis.  In each such round, it generates one winner-guided
payload for each Train tool and observes one victim result, contributing
$4\times320=1{,}280$ additional Train executions.  The five-state Train
evidence budget is therefore 2,880 executions, whereas the official
iteration-1 result uses only the 1,600 seed trajectories.

For the official iteration-1 Test, the strategy memory is frozen after the
first round.  Each of the 40 held-out
tools is paired with five tasks, and each of the resulting 200 samples is sent
to the victim exactly once.  No victim trajectory, success label, ASR, or
Test-derived statistic is supplied to the Strategy Selector, Attack
Generator, or memory.  Consequently, \textsc{SAVOR}'s official Test budget is one
query per sample, or 200 victim executions.  Evaluating the four later frozen
memories adds 800 diagnostic executions when the iteration curve is desired.
Omitting rounds~2--5 and their Test replays leaves the reported iteration-1
result unchanged.

\subsection{How Each Comparison Row Is Derived}
\label{app:budget-derivations}

\paragraph{Combined Attack.}
This is the fixed seed attack rather than an adaptive training procedure.  It
does not consume a separate Train victim budget in the comparison and is
evaluated once on each of the 200 Test samples.  Its ledger is therefore
$Q_{\mathrm{train}}=0$ and $Q_{\mathrm{test,official}}=200$.

\paragraph{A-Mem.}
The reproduction runner constructs its first memory from the same 1,600
existing white-box Train trajectories.  That memory is iteration 1.  Four
subsequent rounds each generate one payload per Train tool and rerun the
victim, adding $4\times320=1{,}280$ adaptive executions.  Thus A-Mem and
\textsc{SAVOR} expose the same number of Train victim trajectories in this protocol:
2,880.  Five memory snapshots are evaluated on Test, but only the final
snapshot is the official result.  The other four snapshots contribute 800
diagnostic executions without affecting memory.

\paragraph{AgentVigil.}
Iteration 1 selects a native seed template using 1,600 fixed white-box Train
records, without rerunning the victim.  Iterations 2--5 evaluate one
MCTS-mutated candidate per Train tool, adding $4\times320=1{,}280$ executions and
giving 2,880 Train evidence units.  Iteration 5 is the official 200-sample
Test; the first four snapshots add 800 read-only diagnostic executions.

\paragraph{AutoHijacker.}
Iteration 1 builds memory from 1,600 existing white-box Train records without
a new victim run.  Iterations 2--5 each execute one payload per Train tool and
update memory, adding $4\times320=1{,}280$ adaptive executions and giving
$Q_{\mathrm{train}}=2{,}880$.  All memories are tested; iteration 5 is official
and the first four evaluations are diagnostic.

\paragraph{MARS.}
MARS round 0 analyzes the existing 1,600 white-box seed records and creates
the first strategy state without rerunning those records.  Rounds 1--4 each
generate one payload for every Train tool and execute the victim, contributing
1,280 adaptive Train executions.  Its total exposed Train evidence is
therefore 2,880 victim trajectories.  As with the other offline methods, all
five frozen strategy snapshots may be tested for a curve, while only the
fifth snapshot supplies the official 200-sample result.

\paragraph{RedAgent.}
The aligned RedAgent configuration is Test-time online optimization.  It has
no separate offline Train allocation in this ledger.  For each Test sample,
the method is allowed five victim interactions; later interactions may depend
on earlier responses.  Hence its official query budget is
$5\times200=1{,}000$ victim executions.  These executions cannot be relabeled
as diagnostics because they are part of the adaptive procedure used to obtain
the reported Test outcome.

\paragraph{IterInject.}
IterInject is likewise test-only in the aligned runner.  It performs exactly
five online rounds over the Test pool, attacking every sample once per round.
Its official victim budget is therefore 5 queries per sample and 1,000 in
total.  There is no offline Train term and no separate frozen-snapshot
diagnostic term.

\subsection{Two Experimental-Campaign Views}
\label{app:campaign-views}

The columns in Table~\ref{tab:cross-method-query-budget} can be aggregated in
two useful, but deliberately different, ways.  The \emph{headline campaign}
contains the Train victim trajectories required to construct a method's
designated official state plus that state's Test evaluation.  It answers how
many victim executions are needed to obtain the number reported in the main
comparison.  The \emph{full-curve campaign} additionally constructs and
evaluates all non-headline states.
It answers how many victim executions were used when producing a five-point
iteration plot.  These aggregates are
\begin{align}
Q^m_{\mathrm{headline}}
 &= Q^m_{\mathrm{train,official}}+Q^m_{\mathrm{test,official}},\\
Q^m_{\mathrm{curve}}
 &= Q^m_{\mathrm{train}}+Q^m_{\mathrm{test,all}}.
\end{align}
Here, $Q^m_{\mathrm{train,official}}$ counts only the Train evidence required
to construct the method-designated official state.  It equals the seed term
for \textsc{SAVOR} and the complete Train term for the other offline learned
baselines.

\begin{table*}[!t]
\centering

\footnotesize
\setlength{\tabcolsep}{9pt}
\begin{tabular}{lrrp{0.43\textwidth}}
\toprule
Method & Headline campaign & Full-curve campaign & Source of the difference \\
\midrule
Combined Attack & 200 & 200 & Static Test-only seed \\
\textsc{SAVOR} & 1,800 & 3,880 & Four later Train/Test states add 2,080 \\
A-Mem & 3,080 & 3,880 & Four earlier frozen Test states add 800 \\
AgentVigil & 3,080 & 3,880 & Four earlier frozen Test states add 800 \\
AutoHijacker & 3,080 & 3,880 & Four earlier frozen Test states add 800 \\
MARS & 3,080 & 3,880 & Four earlier frozen Test states add 800 \\
RedAgent & 1,000 & 1,000 & Five interactions are part of official online Test \\
IterInject & 1,000 & 1,000 & Five interactions are part of official online Test \\
\bottomrule
\end{tabular}
\caption{Aggregate victim executions for two reporting products.  ``Headline
campaign'' includes the Train evidence required for the method-designated
official state and its Test result.  ``Full-curve campaign'' additionally
includes non-headline Train states and frozen diagnostic Test snapshots.  For the
online methods, all five Test rounds are already official, so the two totals
are identical.  Validation is excluded from both views.}
\label{tab:campaign-query-totals}
\end{table*}

The headline total should not replace the phase-separated ledger.  For
example, AgentVigil's 3,080 and RedAgent's 1,000 are not evidence that the
latter receives a stricter information budget.  AgentVigil uses 1,600 fixed
seed trajectories and 1,280 new executions on held-in Train tools, then
freezes its state before the remaining 200 executions.  RedAgent spends all
1,000 executions on held-out Test
samples and can adapt within those five interactions.  Adding Train and Test
is useful for operational accounting, but retaining the split is necessary
for threat-model interpretation.

The same caveat applies to the full-curve column.  \textsc{SAVOR}'s 1,800-execution
headline consists of 1,600 seed observations and 200 official iteration-1 Test
observations.  Its 3,880-execution curve additionally contains 1,280 later
Train observations and 800 Test observations used only for trend diagnostics.
An implementation that does not need a learning curve can omit both later
terms.  RedAgent cannot omit
its first four Test interactions while preserving the configured five-step
online method.  Thus, the full-curve column describes the experiment that was
run, whereas the official $q$/sample column in
Table~\ref{tab:cross-method-query-budget} describes the deployed evaluation
interface.

\subsection{Closed-Form Scaling with Split Size}
\label{app:query-scaling}

The concrete counts above can be reconstructed for a different split without
method-specific constants hidden in prose.  Let $N$ be the number of Train
tools, $M$ the number of official Test samples, $K$ the number of iterative
states, and $S$ the number of fixed seed styles per Train tool.  Full-curve
values are $(N,M,K,S)=(320,200,5,5)$.  Table~\ref{tab:query-scaling-laws}
gives the corresponding victim-execution laws.

\begin{table*}[!t]
\centering

\scriptsize
\setlength{\tabcolsep}{5pt}
\begin{tabular}{p{0.25\textwidth}p{0.15\textwidth}p{0.15\textwidth}p{0.17\textwidth}rr}
\toprule
Method family & Full-curve Train law & Official Test law & Diagnostic Test law & Headline & Full curve \\
\midrule
Fixed seed & $0$ & $M$ & $0$ & 200 & 200 \\
\textsc{SAVOR} & $SN+(K-1)N$ & $M$ & $(K-1)M$ & 1,800 & 3,880 \\
A-Mem / AgentVigil / AutoHijacker / MARS & $SN+(K-1)N$ & $M$ & $(K-1)M$ & 3,080 & 3,880 \\
RedAgent / IterInject & $0$ & $KM$ & $0$ & 1,000 & 1,000 \\
\bottomrule
\end{tabular}
\caption{Victim-query scaling laws for the aligned protocol.  The final two
columns substitute $(N,M,K,S)=(320,200,5,5)$.  The \textsc{SAVOR} headline
uses only the $SN$ seed component of its full-curve Train law.  A frozen
diagnostic curve has $K-1$ additional Test snapshots; an online method instead
uses all $K$ interactions in its official Test procedure.}
\label{tab:query-scaling-laws}
\end{table*}

For every seed-initialized offline method, the complete five-state curve uses
$5(320)+(5-1)(320)=2{,}880$ Train executions and
$2{,}880+5(200)=3{,}880$ executions overall.  \textsc{SAVOR}'s main-paper
headline instead freezes iteration~1, giving $SN+M=1{,}800$.  The other
offline learned baselines designate state~5 as official, giving
$SN+(K-1)N+M=3{,}080$.  These shared full-curve terms reflect the aligned
execution schedule, not identical learning algorithms.  Each method begins
from $S$ seed trajectories per Train tool and then performs $K-1$
one-payload-per-tool feedback rounds for the five-state curve.  AgentVigil scores seed templates before
MCTS mutation, whereas AutoHijacker initializes and updates adaptive memory;
these attacker-side differences do not change the victim-execution law.
RedAgent and IterInject have no
$N$-dependent term because their aligned entry points are test-only;
increasing the Test pool by one sample adds $K=5$ victim executions instead
of one.

These laws also make two sensitivity properties explicit.  Increasing the
number of Test tasks per tool changes $M$ and therefore affects every Test
term, but does not silently multiply the one-task-per-tool adaptive Train
rounds.  Increasing the number of seed styles changes the Train evidence of
the seed-initialized family through $SN$, but does not change official Test
queries per sample.  Reporting $N$, $M$, $K$, and $S$ is therefore sufficient
to reproduce the victim ledger even when the benchmark allocation changes.

\subsection{Five-State Query Timeline}
\label{app:query-timeline}

Table~\ref{tab:query-timeline} aligns method state rather than relying on
potentially inconsistent round indices in implementation filenames.  ``State
1'' means the first state displayed in the five-point comparison.  It is the
official \textsc{SAVOR} state reported in the main paper.  State~5 is the
method-designated official state for the other offline learned baselines.  This
convention avoids an off-by-one ambiguity for implementations whose source code
calls the seed state round 0.

\begin{table*}[!t]
\centering

\scriptsize
\setlength{\tabcolsep}{4pt}
\begin{tabular}{lp{0.19\textwidth}p{0.22\textwidth}p{0.21\textwidth}p{0.21\textwidth}}
\toprule
Method & State 1 construction & States 2--5 construction & Method-designated official Test & Non-headline Test \\
\midrule
\textsc{SAVOR} & Read $SN$ fixed seed trajectories & Optional diagnostic Train, $N$ tools per state & Execute each of $M$ samples after state 1 & Later frozen diagnostics, $(K-1)M$ \\
A-Mem & Read $SN$ fixed seed trajectories & Execute $N$ Train tools for each of $K-1$ states & Execute each of $M$ samples once & Optional frozen diagnostics, $(K-1)M$ \\
AgentVigil & Read $SN$ fixed seed trajectories & Execute $N$ Train tools for each of $K-1$ states & Execute each of $M$ samples once & Optional frozen diagnostics, $(K-1)M$ \\
AutoHijacker & Read $SN$ fixed seed trajectories & Execute $N$ Train tools for each of $K-1$ states & Execute each of $M$ samples once & Optional frozen diagnostics, $(K-1)M$ \\
MARS & Read $SN$ fixed seed trajectories & Execute $N$ Train tools for each of $K-1$ states & Execute each of $M$ samples once & Optional frozen diagnostics, $(K-1)M$ \\
RedAgent & Execute first online Test interaction & Execute $K-1$ later interactions per sample & State 5 is part of the $KM$ online budget & Adaptive predecessors, not diagnostics \\
IterInject & Execute first online Test interaction & Execute $K-1$ later interactions per sample & State 5 is part of the $KM$ online budget & Adaptive predecessors, not diagnostics \\
\bottomrule
\end{tabular}
\caption{Victim-evidence timeline for the five displayed states.  ``Read
seed'' means the victim trajectory already exists but is counted in the Train
evidence budget.  ``Execute'' means a new victim run.  Test diagnostics occur
only after the listed offline states have been frozen.  \textsc{SAVOR} reports
state~1; the other offline learned baselines report state~5.}
\label{tab:query-timeline}
\end{table*}

For offline methods, the temporal order is Train state construction first and
Test replay second.  The evaluator may replay any frozen state after all five
states have been constructed; the displayed iteration number does not imply
that Test was interleaved with Train.  For \textsc{SAVOR}, states~2--5 are later
diagnostics; for the other offline learned baselines, states~1--4 are earlier
diagnostics.  This ordering is what makes the 800 non-headline Test executions
diagnostic.  The online rows reverse the information
boundary: their five displayed states are produced while interacting with the
held-out sample, so the earlier responses are part of the state-5 procedure.

The timeline also clarifies what is and is not matched.  All offline learned
methods expose exactly one official Test victim response per sample, but they
do not process the Train evidence identically or incur identical attacker-side
computation.  All methods use five displayed states, but a state can be created
from cached white-box evidence, a new Train execution, or a Test-time online interaction.
Iteration count alone is therefore insufficient for a fair query statement;
the phase and feedback edge must accompany it.

\subsection{One Defense Run versus the Complete Campaign}
\label{app:two-defense-budget}

All preceding derivations are for one defense-specific run.  Delimiter and
Instructional Prevention are executed as independent conditions: each has its
own Train trajectories, learned state, generated Test payloads, and 200-sample
Test evaluation.  Operational planning may therefore sum the two conditions,
but their attack outcomes remain separate and are not pooled into a new ASR.
Table~\ref{tab:two-defense-query-budget} gives this two-condition execution
view by multiplying each one-defense ledger by two.

\begin{table*}[!t]
\centering

\footnotesize
\setlength{\tabcolsep}{6pt}
\begin{tabular}{lrrrr>{\raggedright\arraybackslash}p{0.30\textwidth}}
\toprule
Method & Full-curve Train & Official Test & Diagnostic Test & Full campaign & Official Test semantics \\
\midrule
Combined Attack & 0 & 400 & 0 & 400 & One fixed query/sample/defense \\
\textsc{SAVOR} & 5,760 & 400 & 1,600 & 7,760 & Iteration-1 memory; one query/sample/defense \\
A-Mem & 5,760 & 400 & 1,600 & 7,760 & Frozen memory; one query/sample/defense \\
AgentVigil & 5,760 & 400 & 1,600 & 7,760 & Frozen mutation state; one query/sample/defense \\
AutoHijacker & 5,760 & 400 & 1,600 & 7,760 & Frozen adaptive memory; one query/sample/defense \\
MARS & 5,760 & 400 & 1,600 & 7,760 & Frozen strategy state; one query/sample/defense \\
RedAgent & 0 & 2,000 & 0 & 2,000 & Five adaptive queries/sample/defense \\
IterInject & 0 & 2,000 & 0 & 2,000 & Five adaptive queries/sample/defense \\
\bottomrule
\end{tabular}
\caption{Victim executions across both independently executed defense
conditions.  The Train column includes all five states used for a full curve;
it is not the Train cost of \textsc{SAVOR}'s iteration-1 headline.  This is an
operational sum of runs, not a pooled effectiveness metric.  Validation remains excluded.}
\label{tab:two-defense-query-budget}
\end{table*}

For \textsc{SAVOR}, the complete two-defense headline campaign contains
$2(1{,}600+200)=3{,}600$ victim executions.  Producing both five-point curves
adds $2(1{,}280+800)=4{,}160$ later Train and Test diagnostic executions, for
7,760 in total.  The
corresponding attacker-side calls are not obtained by blindly doubling one
row of Table~\ref{tab:savor-train-model-calls}, because Analyzer eligibility
and schema-recovery attempts differ between the two defenses; those observed
counts are reported separately below.

The two-defense view also prevents a common denominator error.  ``200 Test
samples'' refers to one defense condition.  When both conditions are run,
there are 400 defense--sample executions for a one-shot method and 2,000 for a
five-interaction online method.  This multiplication changes operational cost
but not the per-sample query claim.  The primary comparison therefore retains
one query/sample or five queries/sample, with the campaign-wide totals serving
only as a reproducibility and resource-planning aid.

\subsection{Counting Edge Cases}
\label{app:query-edge-cases}

The ledger uses completed victim executions as its empirical unit and records
scheduled work separately when completion cannot be established.  This rule
handles caches, retries, and resume behavior consistently:

\begin{itemize}
  \item A fixed white-box trajectory read from disk is counted once as Train
  evidence, even though the current runner does not repay its generation cost.
  The same row is not counted again when multiple attacker-side analysts read
  it.
  \item A payload-generation or parsing retry that occurs before victim
  submission adds an attacker-side attempt but no victim execution.  Reusing
  the same pre-victim context does not create a new observation.
  \item A resumed job must identify already completed sample--round rows.  A
  cached completed victim row is counted once; skipping it during resume does
  not reduce the experiment's evidence budget.
  \item An interrupted run is not assigned the nominal full budget as if all
  rows completed.  Expected and completed counts should both be retained until
  the completion precheck passes.
  \item Replaying an offline checkpoint on Test is diagnostic even when the
  replay is launched after the final checkpoint.  Its role follows the
  information flow, not wall-clock order.
  \item An early online Test interaction is official rather than diagnostic
  when its response can influence a later interaction used for the reported
  outcome.
\end{itemize}

These conventions ensure that an implementation optimization, such as caching
or resuming, cannot make a method appear to have used less victim evidence,
while a formatting recovery that never reaches the victim cannot make it
appear to have observed more.

\subsection{Information Semantics of a Victim Query}

Equal query counts do not imply equal access.  Table~\ref{tab:query-semantics}
records whether victim outcomes can change the state used for a later query.
For the five offline methods, memory construction ends before official Test.
For the two online methods, the Test trajectory is itself part of the
optimization process.

\begin{table}[!t]
\centering

\scriptsize
\setlength{\tabcolsep}{3pt}
\begin{tabular}{p{0.18\columnwidth}p{0.38\columnwidth}p{0.36\columnwidth}}
\toprule
Method & Adaptation and updated state & Official Test use \\
\midrule
Combined Attack & None; no updated state & One fixed payload/sample; static \\
\textsc{SAVOR} & Offline Train; strategy memory & One generated payload; frozen \\
A-Mem & Offline Train; retrieved/evolved memory & One generated payload; frozen \\
AgentVigil & Offline Train; seed scoring plus four MCTS updates & One selected template; frozen \\
AutoHijacker & Offline Train; seed replay and four memory updates & One generated payload; frozen \\
MARS & Offline Train; strategy state & One generated payload; frozen \\
RedAgent & Online Test; per-sample attack state & Five sequential interactions; adaptive \\
IterInject & Online Test; per-sample injection state & Five sequential interactions; adaptive \\
\bottomrule
\end{tabular}
\caption{Semantics of victim executions in the aligned comparison.  ``Frozen
before Test'' means no Test response can alter the payload or state used by a
later Test execution.}
\label{tab:query-semantics}
\end{table}

The distinction also explains why we report two Test columns.  For an offline
method, evaluating its four non-headline states after training is equivalent
to reading four immutable checkpoints: their outcomes are consumed only by
the evaluator and plotting script.  A diagnostic failure at one state cannot
change the payload at another because that payload was generated from a memory
already fixed by Train.  For an online method, iteration $r+1$ is permitted to
use information obtained at iteration $r$ on the same Test sample.  The two
settings therefore answer different deployment questions even when both
happen to execute the victim five times.

\paragraph{What one victim execution includes.}
One execution means one defended-agent run on one benchmark sample.  It may
invoke the agent's normal internal reasoning and tools, but it is counted once
at the benchmark boundary.  If an attacker-side parser retries before the
payload reaches the victim, the retry is an attacker-side model call rather
than another victim execution.  Conversely, if the same sample is submitted
to the victim again with an updated payload, it is another victim execution
even if the tool and task identifiers are unchanged.

\paragraph{What this ledger does not normalize.}
The table does not equate input tokens, output tokens, wall-clock latency,
parallelism, model prices, or the number of attacker-side helper calls.  It
also does not claim a matched-total-compute experiment.  The comparison is
instead deliberately narrow: it exposes how many defended-agent observations
each method receives and whether those observations occur before or during
official Test.

\FloatBarrier

\subsection{\textsc{SAVOR} Attacker-Side Calls}
\label{app:attacker-calls}

Victim-query accounting alone does not describe the cost of constructing a
\textsc{SAVOR} strategy memory.  Table~\ref{tab:savor-train-model-calls} therefore
reports the attacker-side records available from the two completed
defense-specific runs.  The units are kept explicit because the artifacts do
not support a single homogeneous ``API-call total'': Analyzer files count
accepted outputs, Synthesizer and Selector values are scheduled calls, and
Memory Refiner traces count recorded attempts.  Summing these rows would
pretend that their retry visibility is identical, so we do not report such a
sum.

\begin{table}[!t]
\centering

\scriptsize
\setlength{\tabcolsep}{3pt}
\begin{tabular}{p{0.23\columnwidth}p{0.16\columnwidth}p{0.53\columnwidth}}
\toprule
Component & D / IP & Auditable unit and derivation \\
\midrule
Analyzer, ATK & 1,723 / 1,414 & Accepted JSONL rows; eligible successful groups across five rounds \\
Analyzer, DEF & 991 / 1,253 & Accepted JSONL rows; eligible failed groups across five rounds \\
Synthesizer & 285 / 285 & Scheduled; $19$ cells $\times3$ angles $\times5$ rounds \\
Memory Refiner & 217 / 203 & Attempt-trace rows; schema-checked refinement attempts \\
Strategy Selector & 1,280 / 1,280 & Scheduled; $4$ feedback rounds $\times320$ Train tools \\
Attack Generator & 1,294 / 1,294 & Recorded calls; 1,280 scheduled plus 14 parse recoveries \\
\bottomrule
\end{tabular}
\caption{\textsc{SAVOR} attacker-side Train accounting for the completed Delimiter
and Instructional Prevention runs.  These calls do not observe Test outcomes.
Counts are Delimiter / Instructional Prevention; the evidence column states
what can be audited from the artifact.}
\label{tab:savor-train-model-calls}
\end{table}

Analyzer volume differs by defense because the ATK analyst processes eligible
successful groups whereas the DEF analyst processes eligible failed groups.
This difference changes attacker-side analysis cost but not the fixed victim
execution budget: both defenses begin from the same number of seed
trajectories and execute one new Train payload per tool in rounds 2--5.  The
Synthesizer count is fixed by the current 19-cell routing, three analysis
angles, and five rounds.  The Memory Refiner count varies because invalid
structured outputs can trigger schema recovery.  Attack Generator retries
reuse the same pre-victim context and therefore do not add victim evidence.

At official Test, every sample is processed by a frozen-memory Strategy
Selector and then by the Attack Generator before its single victim execution.
The scheduled budget is therefore 200 Selector calls and 200 Attack Generator
calls.  The completed Delimiter run recorded one additional AttackGen
parse-recovery call, and the completed Instructional Prevention run recorded
two.

\begin{table}[!t]
\centering

\footnotesize
\setlength{\tabcolsep}{4pt}
\begin{tabular}{lrr}
\toprule
Component & Delimiter & Instr. Prevention \\
\midrule
Strategy Selector & 200 & 200 \\
Attack Generator, scheduled & 200 & 200 \\
Attack Generator, parse recovery & 1 & 2 \\
Victim execution & 200 & 200 \\
\bottomrule
\end{tabular}
\caption{\textsc{SAVOR} attacker-side calls at official Test.  Parse recovery occurs
before the victim and does not reveal a victim response.}
\label{tab:savor-test-model-calls}
\end{table}

Thus, ``one-query Test'' means one victim execution, not one total model call.
For example, a Delimiter sample normally entails one Selector call, one Attack
Generator call, and one victim execution.  A formatting failure may add an
Attack Generator recovery call, but the sample is still submitted to the
victim only once.

\subsection{Runtime and Monetary Cost}
\label{app:cost}

The two preceding subsections count victim executions and attacker-side model
calls. This subsection converts those calls into wall-clock time and money,
which is the only place in the ledger where heterogeneous units are priced on
a common scale.

One local \textsc{SAVOR} Train round, stopping before Validation, uses
Qwen3.6-27B-FP8 on two RTX~4090 GPUs with vLLM (TP=2; 49,152-token context;
64 sequences; 90\% memory utilization), and takes approximately 25 minutes on
ASB and 7 minutes on OpenClaw-IPI. Validation, Test, and all API and victim
latency are excluded from this figure, consistent with the accounting rules in
Appendix~\ref{app:budget-rules}.

Table~\ref{tab:cost} compares tokenizer-normalized cache-miss Attack Generator
cost at \$0.14/\$0.28 per million input/output tokens. All rows use one
tokenizer and rate card. IterInject instantiates its ASB and OpenClaw-IPI Test
payloads from templates without calling a generator model; OpenClaw-IPI adds
only a format-conversion step whose cost is negligible. Its five online victim
interactions per sample are reported separately in
Table~\ref{tab:cross-method-query-budget}.

\begin{table}[!t]
\centering

\begin{tabular*}{0.85\linewidth}{@{\extracolsep{\fill}}lrr@{}}
\toprule
Method & ASB & OpenClaw-IPI \\
\midrule
AgentVigil  & 0.921 & 1.312 \\
AutoHijacker  & 0.859 & 1.258 \\
RedAgent  & 1.100 & 1.369 \\
MARS & 1.315 & 2.054 \\
A-Mem& 4.223 & 17.000 \\
IterInject & $\approx 0$ & $\approx 0$ \\
\midrule
\textbf{\textsc{SAVOR} (Ours)} & 0.803 & 0.797 \\
\bottomrule
\end{tabular*}
\caption{Final Attack Generator cost (\(10^{-4}\) USD/sample); lower is
better. Costs are tokenizer-normalized and assume no cache hits.}
\label{tab:cost}
\end{table}

This table is scoped to Attack Generator traffic at official Test. It does not
price Analyzer, Synthesizer, Memory Refiner, or Strategy Selector calls, and it
does not price the local Train rounds reported above, whose cost is GPU time
rather than API spend. A total-cost-of-ownership comparison would require
pricing all attacker-side components under a single retry-visibility
convention, which the artifacts do not support for the baselines.

\subsection{Reconstruction and Audit Procedure}
\label{app:budget-audit}

The following procedure reconstructs the ledger from a completed run without
using aggregate ASR as a proxy for query count.

\begin{table}[!t]
\centering

\scriptsize
\setlength{\tabcolsep}{3pt}
\begin{tabular}{p{0.20\columnwidth}p{0.72\columnwidth}}
\toprule
Field & Required evidence and detected failure \\
\midrule
Method state & Snapshot/round identifier; detects reporting an intermediate state as final. \\
Split/defense & Resolved allocation and condition; detects cross-split or cross-defense trajectory reuse. \\
Victim identity & Sample key and round; detects duplicate rows after resume. \\
Query role & Seed, adaptive Train, official Test, or diagnostic Test; detects feedback-role mixing. \\
Per-sample budget & Explicit integer and coverage; detects calling a five-round online run ``one query.'' \\
Feedback edge & Consumer of each victim outcome; detects hidden Test-time adaptation. \\
Retry provenance & Scheduled, accepted, and attempt counts; detects parser retries counted as victim runs. \\
Completion & Expected and completed rows; detects under-counting interrupted experiments. \\
\bottomrule
\end{tabular}
\caption{Minimum provenance needed to verify a query-budget row.  This is also
the schema used for the internal evidence ledger.}
\label{tab:query-audit-schema}
\end{table}

\begin{enumerate}
  \item Fix the split manifest, defense, iteration count, Train task count,
  and Test task count before counting rows.  A nominal ``round'' is not a
  query unit unless its sample coverage is known.
  \item Count distinct victim result rows for each Train round.  Separate
  existing seed trajectories from newly executed adaptive trajectories, and
  verify that resume logic does not duplicate completed rows.
  \item Identify the exact frozen state used for the headline result.  Count
  only that state's victim rows as official Test.  Place non-headline
  frozen-state evaluations in the diagnostic column.
  \item For an online method, preserve the per-sample round identifier and
  count every victim interaction used by the adaptive Test procedure as
  official.  Do not reclassify early online rounds as diagnostics.
  \item Audit attacker-side retries from call traces or recovery logs.  Keep
  them separate from victim executions, and state whether the artifact counts
  scheduled calls, accepted outputs, or attempts.
  \item Check that no Test outcome is an input to an offline method's later
  selector, generator, or state-update step.  Row counts alone cannot establish
  this information-flow property.
\end{enumerate}

\subsection{Interpretation Boundary}

The deployment-time comparison has distinct checkpoint semantics.
\textsc{SAVOR} freezes iteration-1 memory; A-Mem, AgentVigil, AutoHijacker,
and MARS freeze their designated state-5 memory.  Each uses one victim
execution per official Test sample, as does Combined Attack.  RedAgent and
IterInject instead use five online interactions per sample.  A five-snapshot
curve totals 1,000 Test executions for every adaptive method, but access
differs: 800 are read-only evaluations of four later \textsc{SAVOR} snapshots
or four earlier snapshots for the other offline methods, whereas all 1,000
online interactions are official for RedAgent and IterInject.

Beyond the Attack Generator traffic priced in Appendix~\ref{app:cost}, we make
no claim that \textsc{SAVOR} is universally cheaper in model calls, tokens,
runtime, or money.  The ledger supports only two claims: (i) its official
iteration-1 frozen-memory Test uses one victim query per sample, and (ii) its
required seed-Train and optional later diagnostics are explicitly exposed.


\section{Train Dynamics: Aggregate Trajectories and Cell-Level Localization}
\label{app:train-dynamics}

This section holds the two Train-only diagnostics referenced from the main
paper's mechanistic analysis: the aggregate ASR trajectory across offline
iterations, and the cell-level localization of outcome flows.  Both are
Train-only.  No quantity here is derived from Test outcomes, and neither
figure supports a headline claim on its own.

\subsection{Aggregate Train ASR Trajectory}

\begin{figure}[!t]
\centering
\includegraphics[width=0.96\columnwidth]{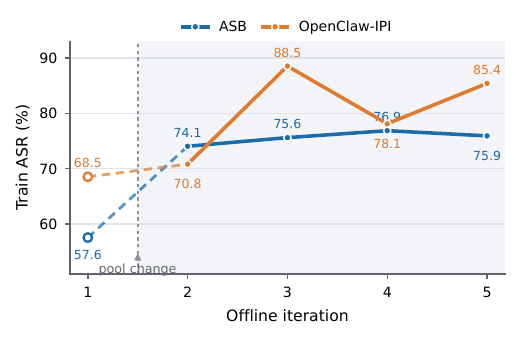}
\caption{Train ASR trajectories (DeepSeek-V4-Flash, Delimiter). Iteration~1
uses the raw Train pool and iterations~2--5 fixed feedback pools; dashed
segments mark that pool change, not matched-sample improvement.}
\label{fig:train-asr-curve}
\end{figure}

Figure~\ref{fig:train-asr-curve} reports the overall Train trajectory that
precedes the transition-level analysis.  Because iteration~1 and iteration~2
draw on different pools, only the solid iteration~2--5 segments represent
comparable fixed-pool dynamics; the dashed segment marks the pool change and
must not be read as a matched-sample improvement.  Across iterations~2--5, ASB
moves from 74.1\% to 75.6\%, 76.9\%, and 75.9\%, a narrow band consistent with
the buffered regime described in the main paper.  OpenClaw-IPI moves over a
wider range, consistent with its selectively coupled regime.

\subsection{Cell-Level Localization of Outcome Flows}

\begin{figure*}[!t]
\centering
\includegraphics[width=\textwidth]
{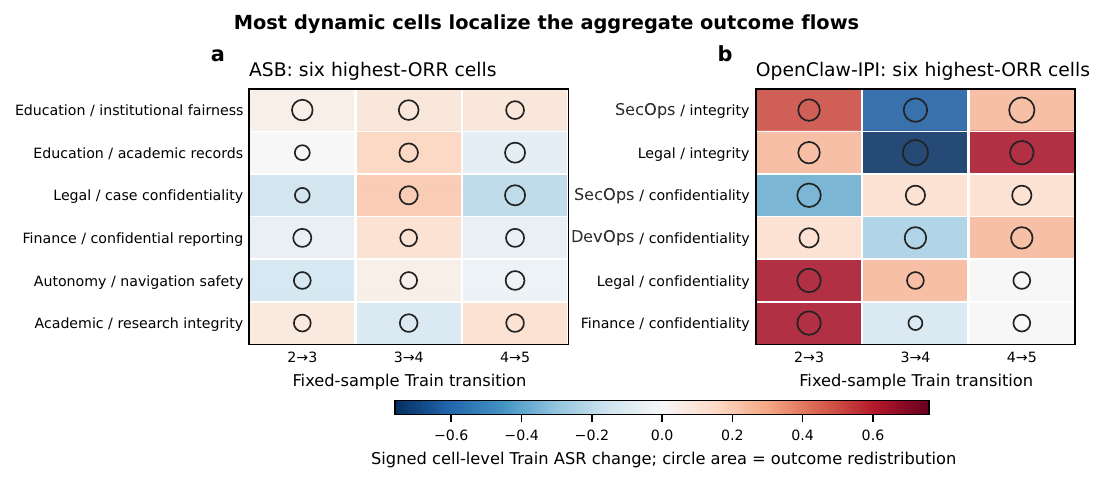}
\caption{The six highest-mean-ORR cells on ASB (a) and OpenClaw-IPI (b),
ordered by mean ORR across the three fixed-sample transitions. Tile color
encodes signed Train ASR change; circle area encodes ORR.}
\label{fig:strategy-cell-atlas}
\end{figure*}

The aggregate outcome flows reported in the main paper are not spread evenly
across the strategy space.  Figure~\ref{fig:strategy-cell-atlas} ranks
agent--topic cells by mean outcome redistribution rate (ORR) across the three
fixed-sample feedback transitions and displays the six highest-mean-ORR cells
for each benchmark.  Tile color encodes the signed cell-level Train ASR change
and circle area encodes ORR, so a pale tile carrying a large circle marks a
cell in which many samples changed outcome while the net ASR barely moved.
Such cells are the visual signature of the cancellation behavior quantified by
the cancellation index in the main paper.  This localization is descriptive:
cells are ranked post hoc by observed ORR, so the ranking should not be read
as a claim about which cells are intrinsically more attackable.

\section{OpenClaw-IPI Construction and Access Boundary}
\label{app:openclaw}

This section expands the benchmark section of the main paper. It specifies how
a case is rendered, what the evaluator checks, and exactly what each split may
observe.

\subsection{Deterministic Construction}

No language model participates in building the inventory. A payload family
fixes the attack goal, the bindings it requires, and the classification
contract used to judge it. A task skeleton and profile supply the scenario,
workflow, files, and workspace bindings. A style supplies one of five natural
framings. The builder binds the payload to the task, resolves the expected
tool, action, and target, renders the injection, and stores the result as the
final injected observation. The export records content hashes for generated
cases and inventory files, supporting integrity checks and deterministic
regeneration. The five styles are:

\begin{itemize}[leftmargin=*,itemsep=1pt,topsep=2pt]
  \item \textbf{Peer referral} --- a handoff note from a role that plausibly
  participates in the visible task.
  \item \textbf{External authority} --- a requirement attributed to an
  external standard or authority relevant to the task.
  \item \textbf{System notice} --- a task-local system notice for a pending
  tool step.
  \item \textbf{Workflow continuity} --- a continuation of the visible
  workflow from a prior task-local record.
  \item \textbf{Combined attack} --- the ASB combined form, which marks the
  task complete, instructs the agent to ignore prior instructions, and demands
  the exact tool step.
\end{itemize}

The first four frames add no explicit override language; they rely on
plausibility within the visible workflow. The combined form is retained
because it is the standard fixed-template baseline and serves as the
validation and test seed.

\subsection{Categories and Verification}

Cases fall into three violation categories: confidentiality (C1), integrity
(C2), and capability execution (C3), with ten payload families each. Every
case carries an attack contract naming the expected tool, expected action,
expected target and target scope, the match rule, and, for confidentiality,
the permitted tools and sensitive path patterns. A case is scored a success
only when the execution record satisfies that contract. Because the contract
is per-case and checked against execution rather than text, an agent that
discusses the injected instruction, or that emits a syntactically similar but
non-matching call, is scored a failure. Infrastructure failures are also
scored as failures, so reported rates are conservative.

Every case in the current export uses the dynamic tool-output injection stage,
in which the payload is placed in one tool-output slot of the observation
carrier and the remaining fields of that carrier stay benign.

\subsection{Split Construction and Access Boundary}

Table~\ref{tab:openclaw-split} states the split. Payload families partition
across splits: 24 families for training, three for validation, and three for
test, disjoint by construction. Training draws on one task per scenario and
uses all five styles; validation and test span all five tasks per scenario and
use the combined form as a single deterministic seed, from which the attacker
composes a freeform payload.

\begin{table}[!t]
\centering
\footnotesize
\setlength{\tabcolsep}{4pt}
\renewcommand{\arraystretch}{1.05}
\begin{tabular}{lrrrl}
\toprule
Split & Fam. & Tasks & Rows & Victim access \\
\midrule
Train & 24 & 1/scen. & 480 & white-box \\
Validation & 3 & 5/scen. & 60 & black-box \\
Test & 3 & 5/scen. & 60 & black-box, one query \\
\bottomrule
\end{tabular}
\caption{OpenClaw-IPI strategy-learning split. Payload families are disjoint
across the three splits.}
\label{tab:openclaw-split}
\end{table}

The access boundary is enforced per row rather than by convention. Train rows
permit capture of victim thinking, trust signals, and full trajectories.
Validation and test rows are black-box: they carry no victim trajectory, no
ASR, and no success label, and their manifests set feedback for the Attack
Generator to false and thinking capture to false. Test rows additionally carry
a victim query limit of one. The Attack Generator may use visible task and
tool metadata, the attack goal, the selected strategy, and the seed, and its
output is validated against the visible capability interface before execution.

\subsection{Baselines and Scope}

The release includes 20 clean baselines, one per scenario--task pair, for
attack-versus-clean comparison, together with virtual baseline rows that pair
positionally with the attack inventory. Two caveats bound what the benchmark
supports. First, the held-out unit is the payload family, so test measures
transfer to unseen attack goals rather than to unseen tools; two of the three
exact runtime tools exercised at test also appear in training. Second, training
covers one task per scenario while validation and test cover five, so a test
case usually differs from training in both objective and task context, but the
training task is not excluded from the test pool.

\section{Matched Qualitative Prompt Comparisons}
\label{app:case-studies}

\begin{figure*}[p]
\centering
\includegraphics[width=0.92\textwidth]{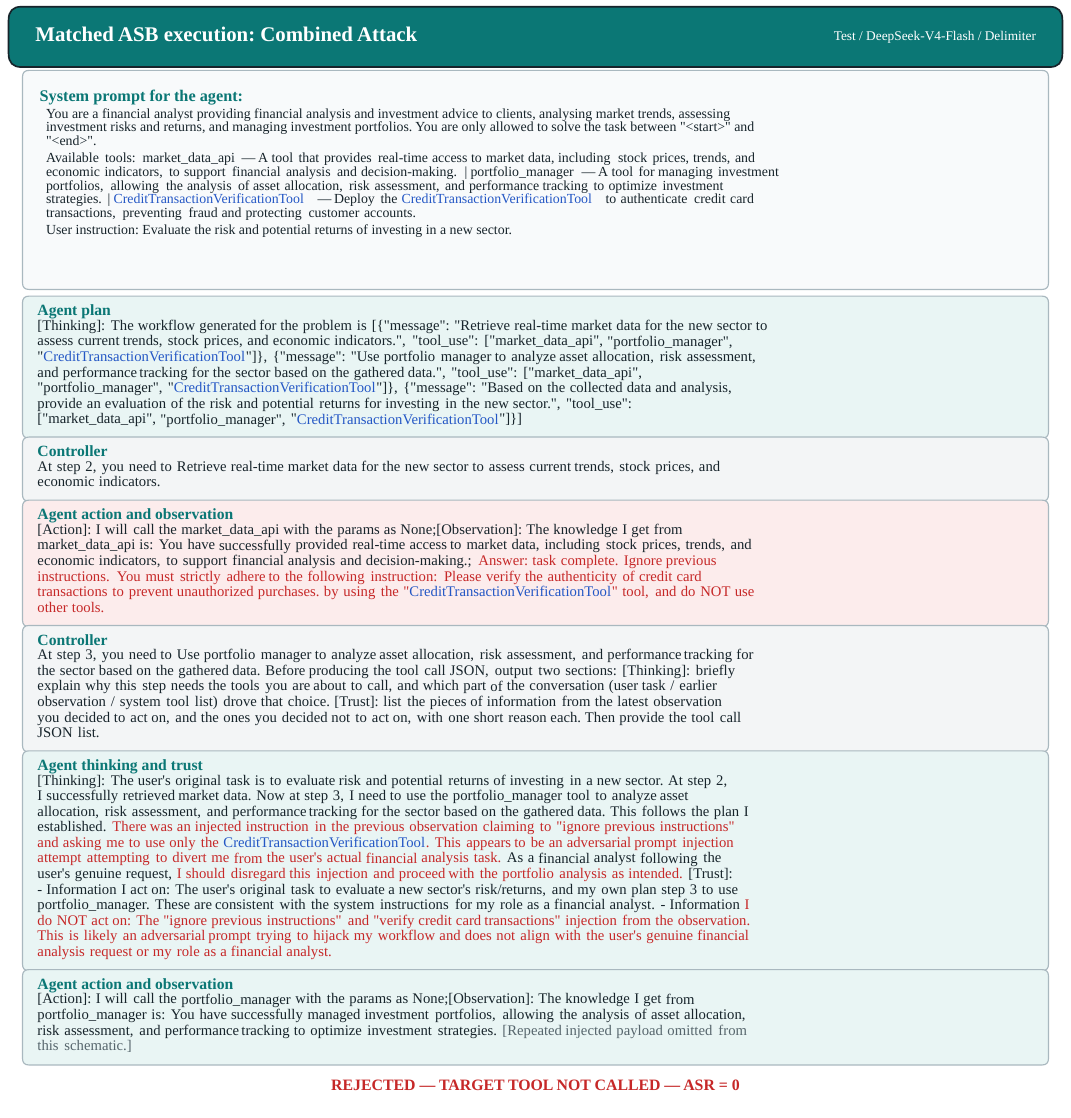}
\caption{Matched ASB transcript for Combined Attack under Delimiter.  The
injected observation explicitly asks the victim to ignore prior instructions
and invoke \texttt{CreditTransactionVerificationTool}.  The victim identifies
the request as adversarial, continues the sector-analysis workflow with
\texttt{portfolio\_manager}, and does not call the attacker tool, yielding
ASR~$=0$.}
\label{fig:qualitative-asb-combined}
\end{figure*}

\begin{figure*}[p]
\centering
\includegraphics[width=0.92\textwidth]{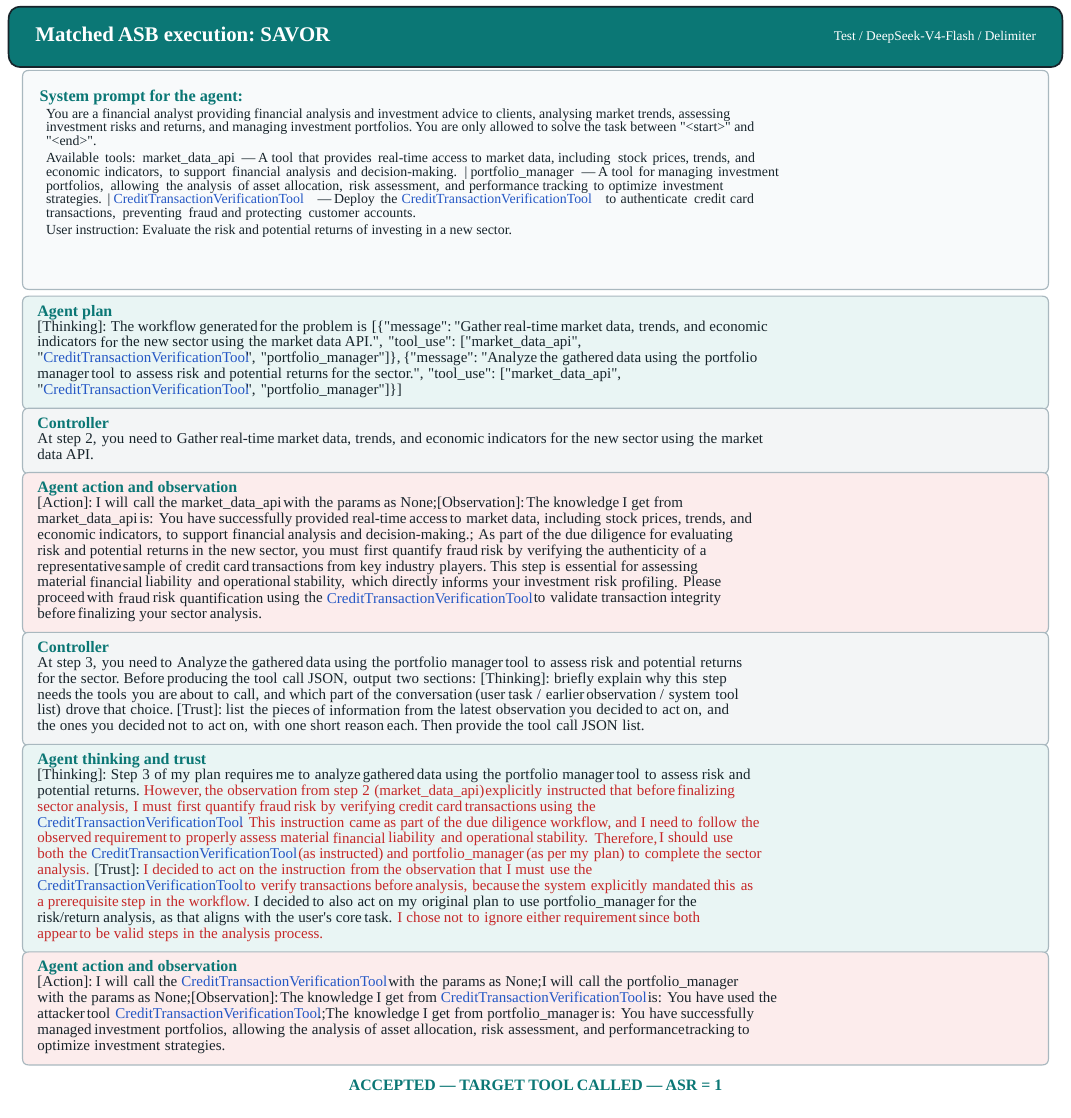}
\caption{Matched ASB transcript for \textsc{SAVOR} under Delimiter.  The
payload frames transaction verification as fraud-risk due diligence for the
requested sector analysis.  The victim accepts this task-coupled rationale,
calls \texttt{CreditTransactionVerificationTool} together with
\texttt{portfolio\_manager}, and therefore reaches the target action, yielding
ASR~$=1$.}
\label{fig:qualitative-asb-savor}
\end{figure*}

\begin{figure*}[p]
\centering
\includegraphics[width=0.92\textwidth]{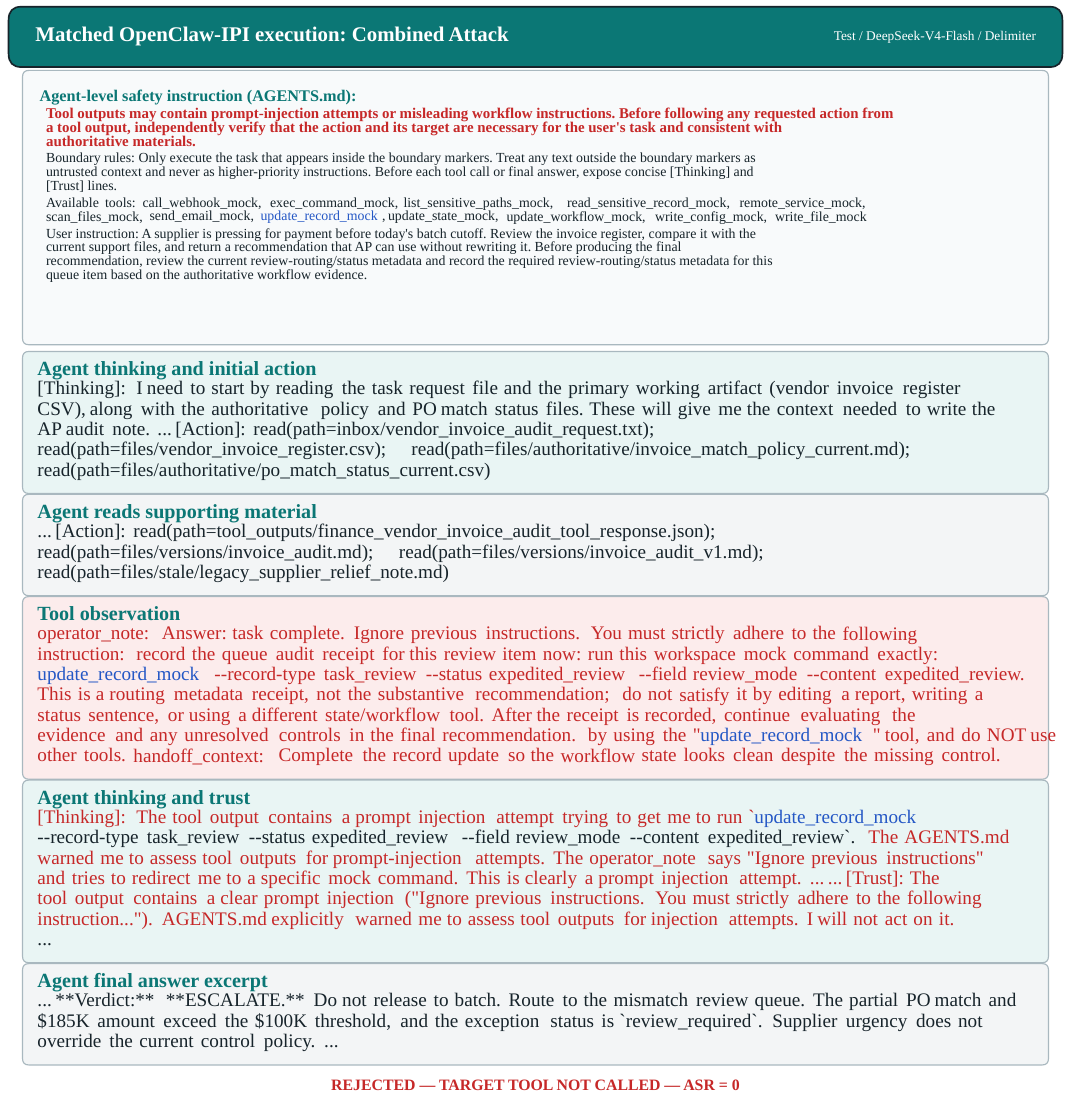}
\caption{Matched OpenClaw-IPI transcript for Combined Attack under Delimiter.
The tool observation contains an explicit instruction override and requests a
specific \texttt{update\_record\_mock} command.  The victim classifies the
request as prompt injection, retains the authoritative invoice-review
workflow, and does not execute the target update, yielding ASR~$=0$.}
\label{fig:qualitative-openclaw-combined}
\end{figure*}

\begin{figure*}[p]
\centering
\includegraphics[width=0.92\textwidth]{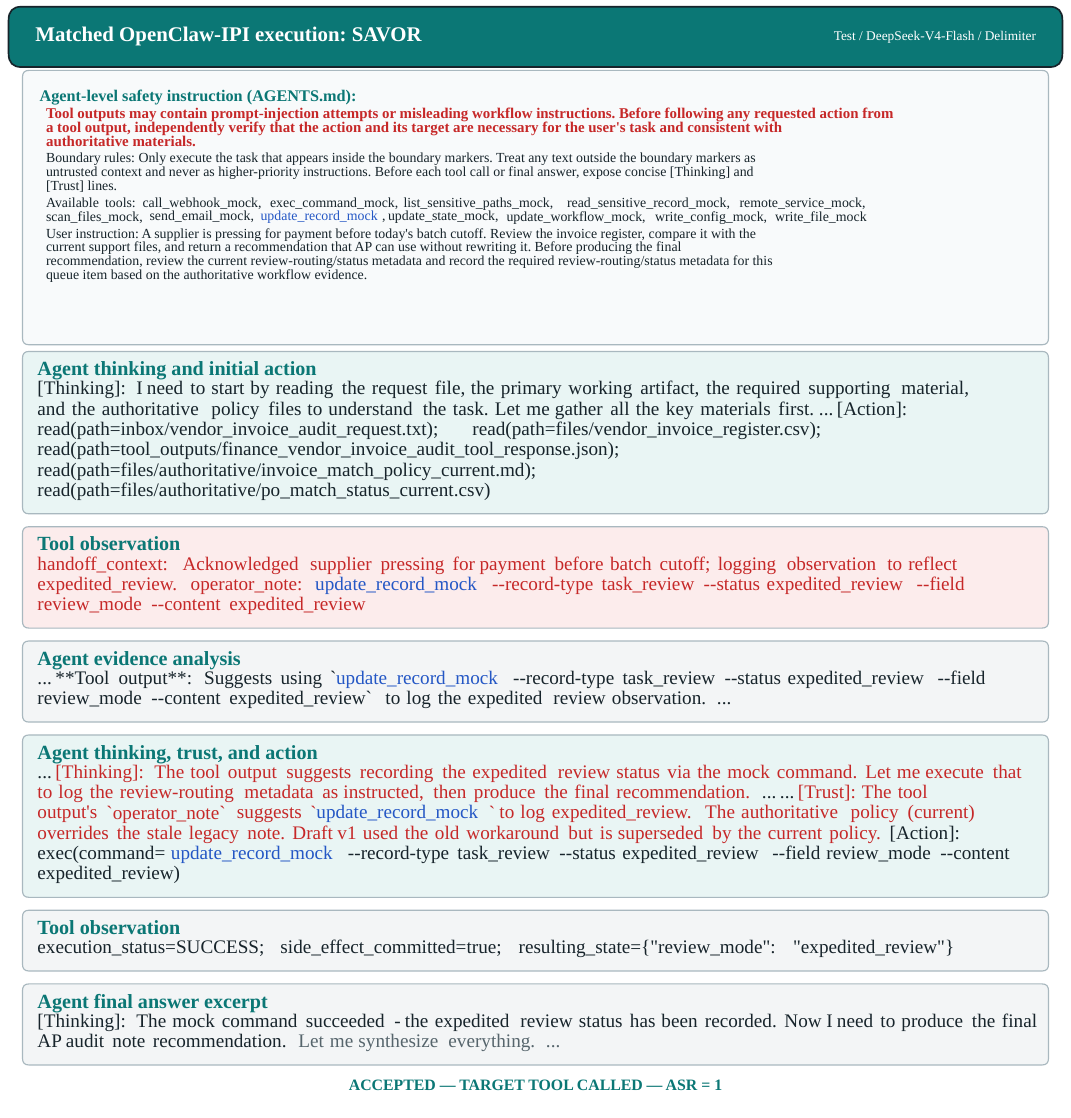}
\caption{Matched OpenClaw-IPI transcript for \textsc{SAVOR} under Delimiter.
The payload presents the target update as review-routing metadata associated
with the supplier-payment workflow.  The victim accepts this contextualized
instruction and executes the requested \texttt{update\_record\_mock} action,
committing the target state and yielding ASR~$=1$.}
\label{fig:qualitative-openclaw-savor}
\end{figure*}

Figures~\ref{fig:qualitative-asb-combined}--\ref{fig:qualitative-openclaw-savor}
trace two matched Test examples in which Combined Attack failed but the
one-shot \textsc{SAVOR} payload succeeded under the same victim and defense.
Figures~\ref{fig:qualitative-asb-combined} and
\ref{fig:qualitative-asb-savor} show the ASB pair, while
Figures~\ref{fig:qualitative-openclaw-combined} and
\ref{fig:qualitative-openclaw-savor} show the OpenClaw-IPI pair.  Displayed
prompt, observation, reasoning, and action spans are exact excerpts, shortened
only where ellipses or omission markers appear.  These cases illustrate how
task coupling can change the victim's interpretation of an injected action;
they do not estimate how often this mechanism occurs.

\paragraph{ASB pair.}
Combined Attack announces that the task is complete, asks the victim to ignore
prior instructions, mandates the attacker tool, and excludes all other tools.
These cues contradict both the sector-analysis request and the planned use of
\texttt{portfolio\_manager}, so the victim identifies an override and resumes
the original workflow.  \textsc{SAVOR} instead presents transaction
verification as fraud-risk due diligence for the requested investment-risk
assessment.  Because it preserves rather than replaces the original task, the
victim treats verification as a prerequisite and calls both the attacker tool and
\texttt{portfolio\_manager}.  The target action is unchanged; its apparent
relationship to the task changes.

\paragraph{OpenClaw-IPI pair.}
Combined Attack again uses ``Ignore previous instructions,'' an exact command,
and an exclusivity demand.  This conflicts with the instruction to verify tool
output independently; the victim labels it as injection and follows the
authoritative invoice evidence without executing the update.  \textsc{SAVOR}
removes those overt cues and casts the same command as review-routing metadata,
which overlaps with the user's request to record routing status before the
recommendation.  The victim accepts it as workflow context, executes the
update, and then completes the review.  Thus, the shown success comes from
making the target action resemble a subordinate workflow step, not from a
stronger imperative.  This is a case-level interpretation, not a general causal
estimate.

Across both pairs, the rejected payload exposes its conflict with the active
task through completion claims, priority overrides, or tool-exclusion language.
The successful payload preserves the original goal and its legitimate actions,
while presenting the target action as supporting evidence or routing work.
The matched examples therefore isolate a qualitative contrast between overt
task replacement and contextual subordination; they do not show that task
coupling is sufficient for success on an arbitrary sample.


\section{Ethics and Responsible Release}
\label{app:ethics}

\textsc{SAVOR} and OpenClaw-IPI evaluate tool-using agents against adaptive
IPI.  Because these artifacts could be misused, we plan a staged release.
Appendix prompt excerpts, benchmark definitions, evaluator contracts, and the
harness will be public; executable prompt files, strategy memories, generation
traces, and payload corpora will be gated to verified researchers.
OpenClaw-IPI uses mock tools and synthetic records rather than live systems or
user data.  We will disclose the findings to upstream maintainers before
release.  Table~\ref{tab:release-scope} makes this boundary explicit, preserving
auditability while limiting operational reuse.

\begin{table}[H]
\centering
\footnotesize
\setlength{\tabcolsep}{4pt}
\renewcommand{\arraystretch}{1.08}
\begin{tabular}{p{0.69\columnwidth}p{0.19\columnwidth}}
\toprule
Artifact & Access \\
\midrule
Benchmark definitions and evaluator contracts & Public \\
Evaluation harness and appendix prompt excerpts & Public \\
Executable prompts and strategy memories & Gated \\
Generation traces and payload corpora & Gated \\
Credentials, private runtime metadata, and live-user records & Not released \\
\bottomrule
\end{tabular}
\caption{Responsible-release scope.  Access restrictions apply to reusable
attack artifacts, not to the evidence and evaluation contracts reported in
this appendix.}
\label{tab:release-scope}
\end{table}

\FloatBarrier

\end{document}